\documentclass{aa}  

\usepackage{graphicx}
\usepackage{natbib}
\usepackage{txfonts}
\usepackage[dvipsnames]{xcolor}
\usepackage[]{hyperref}
\begin{document}

   \title{BLAZ4R and the eROSITA view of $z>4$ blazars}

   \author{T. Sbarrato\inst{1}
          \and
          S. Belladitta\inst{2,3}
          \and
          J. Wolf\inst{2}
          \and
          P. Baldini\inst{4}
          \and
          D. Tubín-Arenas\inst{5}
          \and
          M. Salvato\inst{4}
          \and
          E. Momjian\inst{6}
          \and
          S. H\"ammerich\inst{7}
          \and
          A. Merloni\inst{4}
          \and
          W. Collmar\inst{4}
          \and
          J. Wilms\inst{7},
          }

   \institute{INAF -- Osservatorio Astronomico di Brera, Via Bianchi 46, I-23807 Merate, Italy\\
              \email{tullia.sbarrato@inaf.it}
              \and
              Max-Planck-Institut f{\"u}r Astronomie, K{\"o}nigstuhl 17, 69117 Heidelberg, Germany \\
              \email{belladitta@mpia.de}
              \and
              INAF -- Osservatorio di Astrofisica e Scienza dello Spazio, via Gobetti 93/3, 40129, Bologna, Italy
              \and
              Max-Planck-Institut für extraterrestrische Physik, Gießenbachstraße 1, 85748 Garching, Germany
              \and
              Leibniz-Institut für Astrophysik Potsdam, An der Sternwarte~16, 14482 Potsdam, Germany
              \and
              National Radio Astronomy Observatory, P.O. Box O, Socorro, NM, USA
              \and
              Dr.\ Karl Remeis-Sternwarte and Erlangen Centre for Astroparticle Physics, Friedrich-Alexander Universit\"at Erlangen-N\"urnberg, Sternwartstr.~7, 96049 Bamberg, Germany
            }

   \date{ }

  \abstract
   {
   We present  BLAZ4R, the first living catalog of confirmed $z>4$ blazars, with a focus on the contribution of eROSITA, on board of the Spectrum Roentgen Gamma (SRG) spacecraft. 
   Blazars at $z>4$ are rare but powerful probes of how active supermassive black holes evolve in connection to relativistic jets, in the first 2 billion years of cosmic history.
   At these redshifts, X-ray observations are essential for constraining blazars jet power and orientation, enabling effective trace of their parent population. 
   The all-sky surveys 
   provided by eROSITA ensure X-ray detection for  BLAZ4R sources and, in some cases, allow spectral and temporal studies of their jetted emission.
    BLAZ4R includes 54 confirmed blazars, characterized through their X-ray properties, radio spectra and morphology, and multiwavelength profiles. 
   We confirm that jetted sources are significantly more numerous relative to non-jetted counterparts at high-$z$, and that blazars (and therefore the overall jetted population) do not exhibit significantly different features compared to the total active galactic nuclei population in the early Universe. 
   Fast accretion processes that involve relativistic jets are clearly required to justify the existence of fully formed jetted AGN at $z>4$. However, the diverse multiwavelength properties characterizing  BLAZ4R do not yet allow us to identify the specific signatures of these processes. 
   We will continue updating  BLAZ4R to search for such signatures and ultimately understand the early formation of jetted AGN.
   }

   \keywords{quasars -- jets -- X-rays -- high-z 
               }

   \maketitle

\section{Introduction}

Blazars are active galactic nuclei (AGN) characterized by relativistic jets oriented at small angles to our line of sight (e.g., \citealt{blandford1978,urrypadovani1995}). 
Due to relativistic beaming, blazars are the most luminous persistent sources in the Universe, dominating the high-energy sky at high Galactic latitudes. 
They exhibit strong variability across the entire electromagnetic spectrum, on timescales ranging from minutes to years (e.g., \citealt{ciaramella2004,rajput2020,janssen2023,gokus2024}).
 
Their multiwaveband spectral energy distributions (SEDs) well covered from radio to high-energy (X-ray, $\gamma$-ray) typically show two bumps \citep[e.g.\ ][]{impey88,maraschi94,fossati98,ghisellini17}.
The first bump, peaking at lower energies (infrared to X-ray), is usually attributed to synchrotron emission, while the second peak, at higher energies (X-ray to $\gamma$-ray) originates from inverse Compton (IC) processes or from radiation caused by hadronic particles \citep[][]{boettcher07,ghisellini09}.  
High-energy detections guarantee a prominent IC radiation, signature of relativistically beamed emission from a jet aligned close to our line of sight. 
Despite ongoing debate about the emitting population, blazars exhibit a well-established and self-similar broad-band spectral profile, with $\gamma$-ray emission serving as the definitive signature for blazar classification at low redshift.

At $z>4$ $\gamma$-ray detection is not efficient: the sensitivity of the Large Area Telescope (LAT) onboard the {\it Fermi} satellite \citep{atwood09} allows only for the observation of few blazars in this redshift range \citep{ackermann17,liao18,kreter20}.
X-ray observations are therefore crucial for constraining IC dominance over the synchrotron component: intense, hard X-ray emission that is significantly brighter than the radio and NIR/optical emission indicates strong IC contribution.
Up to now, the X-ray detection of $z<4$ blazars was performed after dedicated observations of good blazar candidates, but the X-ray all-sky survey of the extended ROentgen Survey with an Imaging Telescope Array \citep[eROSITA][]{predehl21} onboard the Spectrum-RG
mission \citep{sunyaev21} can make a difference in their study also at higher redshift. 
\cite{haemmerich25} demonstrated the efficiency of this instrument in surveying the sky and producing a blazar all-sky survey in the soft X-rays, with impressive detection efficiency already within .
the first eROSITA All-Sky Survey (eRASS1, \citealt{merloni2024}).
eROSITA will thus play a significant role in the blooming field of high-$z$ blazar search and characterization. 
In this work, we investigate what constraints this instrument places on the known $z>4$ blazar population.

High-$z$ blazars are key tracers of the overall jetted population in the first 2 billion years of the Universe. 
Because of their peculiar orientation, we can assume the existence of a significant parent population for every blazar observed, i.e.\ for each aligned jet there must exist a number of sources consistent in mass, accretion, jet and host features, in the same redshift bin, with their jets oriented randomly in the sky.
If strictly defined, i.e.\ when observed with a viewing angle $\theta_{\rm v}$ smaller than their jet beaming angle $\theta_{\rm b}\simeq1/\Gamma$ ($\Gamma\sim8-15$ being the bulk Lorentz factor of the emitting region), their parent population is inferred to be $2\Gamma^2\sim130-450$ per blazar observed \citep{volonteri11,ghisellini10,sbarrato12,sbarrato21,belladitta20}.
Blazars are extremely efficient  instruments to study the population of sometimes elusive high-$z$ jetted AGN, and our up to now sparse sampling of such sources is currently reaching large enough numbers to begin population studies. 

The importance of studying AGN in the first two billion years of the Universe resides in the so-called ``mass problem" of the earliest active supermassive black holes (SMBHs). 
The more than 500 quasars discovered up to now at very high redshift appear to be significantly overmassive with respect to the time available for their evolution \citep[see][and references therein]{fan23}: even assuming continuous accretion at the Eddington limit, massive black hole seeds are needed to justify masses of $\sim10^9M_\odot$ at $z\sim4-5$, and often not sufficient. 

The $z>4$ jetted fraction appears to be larger than what observed at lower redshift \citep[][]{sbarrato21,capetti24}.
At first this was rightfully considered a further obstacle to the fast formation and accretion of early SMBHs, since relativistic jets are generally associated to maximally spinning black holes, that in turn have more radiatively efficient accretion disk and thus a less efficient actual mass accretion on the central engine. 
The coexistence of an excess of relativistic jets and extreme SMBH masses in the first 1.5-2 billion years of the Universe has suggested that jets might actually play an active role in the fast assembly of these sources.
Despite their possible causal relation being unclear, various options have been explored in the literature: jets might be triggered by or after super-critical events \citep{alzati25}, they might  facilitate fast accretion by exploiting part of the released gravitational energy and allowing for a faster motion of matter in the disk compared to equivalently bright non-jetted quasars \citep{jolley08,ghisellini13}, or they might affect the influx of matter toward the nuclear region, supplying a larger amount of material that would allow a faster SMBH accretion. 

$z>4$ blazars are thus a clear asset in our knowledge of the early formation and evolution of massive black holes and their environment, providing a privileged point of view on their jet features and the accretion system powering them, and playing a statistically relevant role in tracing the jetted population. 
In this work we introduce  BLAZ4R, the first catalog of confirmed $z>4$ blazars, that will be kept up to date with future newly classified blazars in this redshift range.
BLAZ4R is designed to monitor the evolution of the known $z>4$ blazar population, with eROSITA facilitating the characterization of their jet power and rough orientation.

In this paper we present the  BLAZ4R starting sample selection (Section \ref{Sec:sample}), and the collection of multiwavelength data, with specific focus on eROSITA and radio catalogs (Section \ref{Sec:data}). We then characterize the sample in terms of X-ray variability, broad-band SED studies, and radio features (Section \ref{Sec:character}).
After discussing peculiar sources (Section \ref{sec:notes}), we explore BLAZ4R in the context of high-$z$ jetted vs.\ non-jetted populations, and their accretion features (Section \ref{Sec:population}).

In the following we adopt a flat cosmology with a Hubble parameter $H_0$ = 70 km s$^{-1}$ Mpc$^{-1}$, $\Omega_m$ = 0.30, and $\Omega_{\Lambda}$ = 0.70.
Radio spectral indices are given assuming S$_{\nu} \propto \nu^{-\alpha}$.

\section{Sample}
\label{Sec:sample}

This work aims at collecting and analysing the most complete sample up to date of known  $z>4$ blazars.
To do so, we looked at all the confirmed blazars available in the literature.
We consider all classification methods, that articulate in radio based classifications \citep[flat radio spectrum, Doppler boosting and short scale variability, e.g.\ ][]{caccianiga19,caccianiga24,coppejans16,krezinger2022,banados25,krezinger26}, optical to X-ray features comparison \citep{ighina19}, broad band SED analysis \citep[e.g.\ ][]{belladitta20,ghisellini15,sbarrato12,sbarrato15}, and a serendipitous high-energy detection \citep{romani04}, to collect all known blazars in this redshift bin.
We obtained a sample of 53 sources, listed in Table \ref{Table:sample} with their specific references. 

These 53 sources are the starting point of  BLAZ4R, a living catalog of $z>4$ blazars that will be updated as long as new objects will be discovered and classified. 
For all new sources, we will repeat the broad band analysis explained in this work: retrieve public multiwavelength data, model their SEDs, and check for radio peculiarities. 
Specific focus will be put on their nuclear features (SMBH mass and disc luminosity) and IC component, in order to have enough information on their accretion regime and jet orientation. The latter is crucial for a proper blazar classification.
All info we will derive in this work, i.e.\ mass, accretion luminosity, parameters describing the phenomenological SED, SED images and other possible notes will be available on the dedicated website \url{https://blaz4r.brera.inaf.it/} . 
The  BLAZ4R catalog and website will be continuously updated, providing a consistent reference for the known jetted AGN pointed toward us. 

We add to our core sample 11 $z>4$ radio bright quasars, whose radio features are not considered conclusive in terms of their jet orientation and thus specific classification, and are lacking X-ray observations to constrain the beaming and orientation. 
We introduce these sources in our starting sample as ``candidates", with the possibility of having them classified thanks to eROSITA observations.
They are also included in Table \ref{Table:sample}, labeled accordingly and with their references. 

We cross-correlated the sample with eRASS:5 scans, i.e.\ with the four completed scans and the available 5th scan. All the cross-correlation details are reported in Section \ref{sec:cross}. 
17 out of the 53 blazars are detected by eROSITA.
Of the 11 candidates, J020228-170827 is clearly detected in eRASS:5, and its detection and subsequent broad band SED modeling allowed us to classify it as a new $z\simeq5.6$ blazar.
 BLAZ4R final sample is thus composed of 54 $z>4$ confirmed blazars.
18 of these sources are part of the  BLAZ4R-eRO sample, i.e.\ all $z>4$ blazars detected in eRASS:5 data.

\subsection{eRASS:5 cross-correlation and association statistics}
\label{sec:cross}

We have identified eROSITA X-ray counterparts to the starting sample, consisting of 52  BLAZ4R sources and 11 radio-bright quasars, by cross-matching it with the eRASS1 to eRASS5 single-scan hemisphere catalogs, as well as with the cumulative eRASS:5 catalog \citep[for details on the first-pass survey, see][]{merloni2024}. The individual eRASS catalogs were initially matched to Data Release 10 of the Dark Energy Survey Instrument (DESI) Legacy Imaging Survey (hereafter LS10), CatWISE2020 \citep{marocco21}, and the third \textit{Gaia} Data Release \citep{gaia23}, following the methodology presented in \cite{salvato25}. This approach uses astrometry and random-forest--trained photometric priors within the Bayesian cross-matching framework \textsc{NWAY} \citep{salvato18,salvato22}. \citet{salvato22} report a 94\% accuracy for this statistical matching methodology based on tests using \textit{Chandra} control samples.

\begin{figure}
    \centering
    \includegraphics[width=\linewidth]{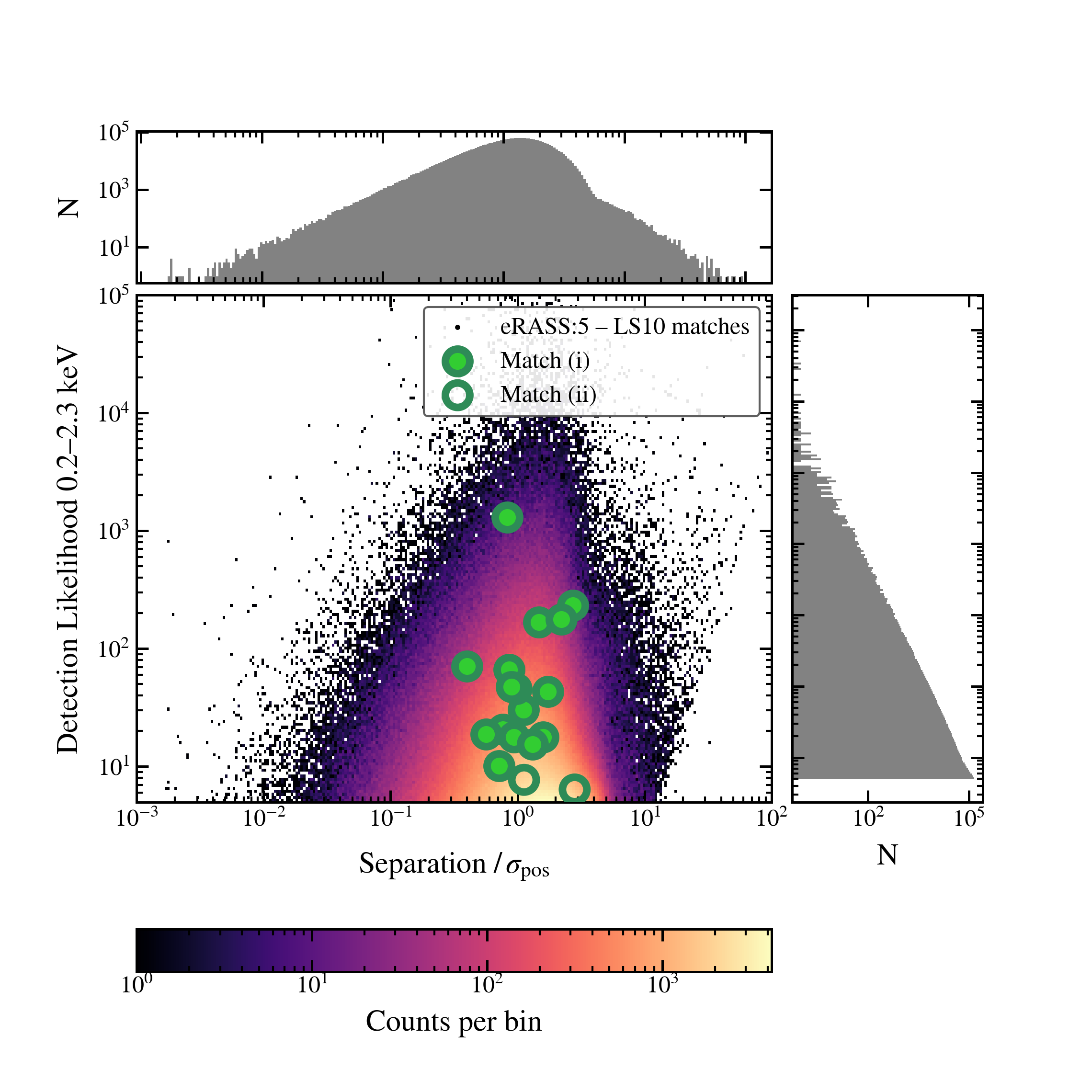}
    \caption{Detection likelihood of blazar-associated eRASS sources from the eRASS:5 catalogue as a function of the angular separation between the X-ray source and the blazar optical position, normalized by the X-ray positional uncertainty. The background density shows the distribution of all eRASS:5–LS10 matches, while highlighted points indicate the blazar associations. Full circles denote blazars matched directly to the LS10 counterpart of an eRASS source. Empty points show blazars matched to an eRASS source within 15\farcs. Blazar counterparts preferentially occur at higher detection likelihoods compared to typical LS10 matches, while spanning a similar range in normalized separation.}
    \label{fig:matches}
\end{figure}

We apply a two-step matching approach: (i) we first match the optical positions of our starting sample to LS10 optical counterparts of eRASS sources \citep[see][for an extensive discussion on the eROSITA LS10 catalog in relation to blazars, their association and completeness]{haemmerich25}; (ii) we then match the optical positions of our starting sample to any eRASS X-ray positions within $15\farcs$, regardless of the position of the best-matching LS10 counterpart. We obtain 16 matches from step (i) and two additional matches (J014132--542749 and J081333+350810) from step (ii). J014132--542749 was missed in step (i) because the \textsc{NWAY} cross-match identified another LS10 counterpart at a larger distance from the X-ray centroid ($14\farcs34$) than the blazar, owing to its more favorable photometric properties. J081333+350810 was missed in step (i) because the source does not have an LS10 counterpart candidate. We note that J014132--542749 and J081333+350810 are the only two sources with detection likelihood $\mathrm{DET\_LIKE}<10$\footnote{$\mathrm{DET\_LIKE}=-\ln(P)$, where $P$ is the null-hypothesis probability that the observed X-ray signal arises from background fluctuations alone.}. We summarize the matching statistics in Fig. \ref{fig:matches}, where we show the detection likelihood of the blazar-associated eRASS sources (from the eRASS:4.5 catalogue) as a function of the X-ray-blazar separation  normalized by the X-ray positional uncertainty. In addition we show the distribution for the eRASS:5 - LS10 matches. The blazar locus is located at higher detection likelihoods than typical LS10 counterparts but span a similar normalizaed separation range.

17 out of the 18 matches are known blazars from our core  while one came from the radio-bright quasar sample.
The broad band SED modeling (Section \ref{sec:SED} allows us to classify this latter source as a blazar, and include it in  BLAZ4R. 
   \begin{figure}
   \centering
   \vskip -0.8cm
   \includegraphics[width=1.05\hsize]{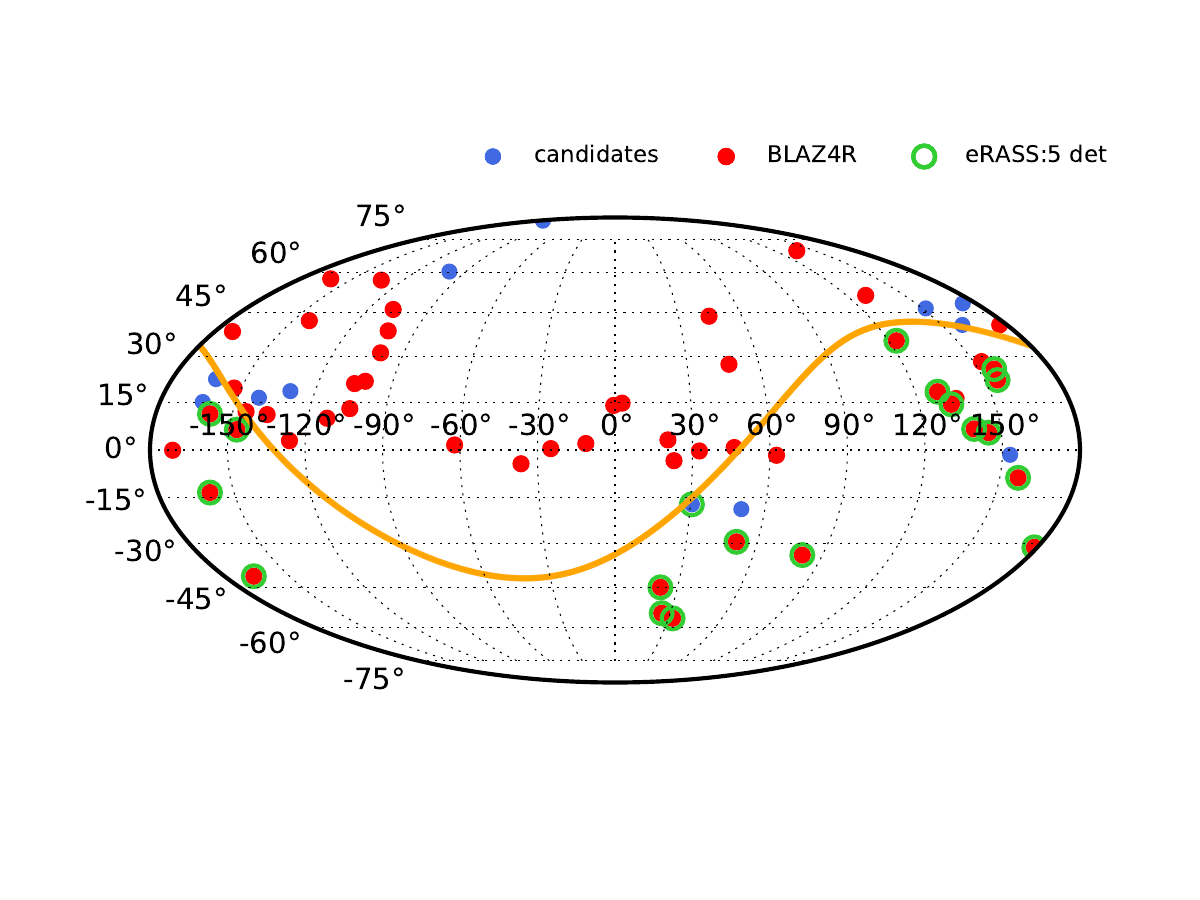}
   \vskip -1.5cm
      \caption{Sky map of the total sample, including the eROSITA footprint limit (yellow line).
               Filled red and blue points show the sources originally classified as confirmed and candidate blazars, respectively. 
               Green circles highlights sources detected in eRASS:5 scans.
               In the eROSITA-DE footprint, only 4 previously classified blazars are undetected. 
              }
         \label{Fig:sky-map}
   \end{figure}
%

%
\begin{table*}
\renewcommand{\arraystretch}{1.1} 
\caption{\tiny Starting sample. 
         References: 
         $^1$\cite{banados23},
         $^2$\cite{banados25},
         $^3$\cite{belladitta19},
         $^5$\cite{belladitta20},
         $^6$\cite{belladitta23},
         $^7$\cite{belladitta25},
         $^8$\cite{caccianiga19},
         $^9$\cite{caccianiga24},
         $^{10}$\cite{coppejans16},
         $^{11}$\cite{ghisellini14},
         $^{12}$\cite{ghisellini15},
         $^{13}$\cite{ghisellini15b},
         $^{14}$\cite{gloudemans22},
         $^{15}$\cite{healey08},
         $^{16}$\cite{ighina19},
         $^{17}$\cite{ighina24},
         $^{18}$\cite{ighina25},
         $^{19}$\cite{khorunzhev21},
         $^{20}$\cite{massaro09},
         $^{21}$\cite{massaro15},
         $^{22}$\cite{romani04},
         $^{23}$\cite{sbarrato12},
         $^{24}$\cite{sbarrato13a},
         $^{25}$\cite{sbarrato13b},
         $^{26}$\cite{sbarrato15},
         $^{27}$\cite{sbarrato22},
         $^{28}$\cite{shemmer06},      
         $^{29}$\cite{sowardsemmerd2003},
         $^{30}$\cite{tao23},
         $^{31}$\cite{wolf24},
         $^{32}$\cite{worsley04},
         $^{33}$\cite{yuan00},
         $^{34}$\cite{yuan03}.
        Objects marked with an $l$ have radio spectral indices taken from the literature. }         
\label{Table:sample}      
\centering  
\tiny
\begin{tabular}{l l l l c l l}
\hline \hline
  Name & $z$ & RA  & Dec  & blazar class. & eRO cts & $F_{[0.2-2.3 {\rm keV}]}$ (erg/cm$^2$/s) \\
\hline
  J001115+144601 & 4.96 & 00:11:15.24 & +14:46:01.8 & y $^{28}$ &  &  \\
  J012202+030951 & 4.0 & 01:22:01.90 & +03:10:02.41 & y $^{29}$ &  & \\
  J012714-445445$^l$ & 4.96 & 01:27:14.32 & -44:54:45.17 & y $^{18}$ & 43.2 & $6.88_{-1.15}^{+1.28}\times10^{-14}$ \\
  J013127-032100 & 5.18 & 01:31:27.34 & -03:21:00.2 & y $^{12}$ &  & \\
  J014132-542749$^l$ & 5.0 & 01:41:32.4 & -54:27:49.9 & y $^3$ & 8.65 & $1.24_{-0.47}^{+0.60}\times10^{-14}$  \\ 
  J020228-170827 & 5.57 & 02:02:28.53 & -17:08:27.75 & c $^{17}$, y $^{\rm here}$ & 16.67 & $2.10_{-0.56}^{+0.64}\times10^{-14}$ \\
  J020916-562650$^l$ & 5.606 & 02:09:16.93 & -56:26:50.44 & y $^{17,31}$ & 21.45 & $2.81_{-0.65}^{+0.74}\times10^{-14}$\\
  J021043-001818 & 4.77 & 02:10:43.16 & -00:18:18.4 & y $^{10}$ &  & \\
  J025758+433837 & 4.07 & 02:57:59.08 & +43:38:37.71 & y $^{8,16}$ &  & \\
  J030437+004653 & 4.305 & 03:04:37.21 & +00:46:53.5 & c $^{27}$ &  &  \\
  J030947+271757$^l$ & 6.1 & 03:09:47.49 & +27:17:57.31 & y $^5$ &  & \\
  J032214-184117$^l$ & 6.09 & 03:22:14.55 & -18:41:17.48 & y $^{18}$ &   & \\
  J032444-291821 & 4.63 & 03:24:44.29 & -29:18:21.22 & y $^{20}$ & 88.01 & $8.60_{-0.98}^{+1.04}\times10^{-14}$ \\
  J041009-013919$^l$ & 6.995 & 04:10:09.05 & -01:39:19.88 & y $^2$ &  & $<1.90\times10^{-14}$\\
  J052506-334305 & 4.383 & 05:25:06.18 & -33:43:05.5 & y $^{32}$ & 392.60 & $4.34_{-0.23}^{+0.23}\times10^{-13}$ \\
  J081333+350810 & 4.922 & 08:13:33.32 & +35:08:10.8 & y $^{27}$ & 7.81 & $2.27_{-0.87}^{+1.06}\times10^{-14}$ \\
  J083548+182519 & 4.41 & 08:35:49.42 & +18:25:20.09 & y $^{8,16}$ & 33.09 & $1.06_{-0.19}^{+0.21}\times10^{-13}$ \\
  J083946+511202 & 4.4 & 08:39:46.22 & +51:12:02.8 & y $^{24}$ &  & \\
  J085111+142337 & 4.307 & 08:51:11.59 & +14:23:37.7 & y $^{27}$ & 29.68 & $8.96_{-1.68}^{+1.94}\times10^{-14}$ \\
  J090132+161506 & 5.63 & 09:01:32.65 & +16:15:06.84 & y $^9$ &  & $<5.80\times10^{-14}$ \\
  J090630+693030$^l$ & 5.47 & 09:06:30.75 & +69:30:30.80 & y $^{22}$ &  & \\
  J091824+063653 & 4.22 & 09:18:24.38 & +06:36:53.3 & y $^{21}$ & 13.23 & $4.24_{-1.20}^{+1.43}\times10^{-14}$ \\
  J094004+052630 & 4.5 & 09:40:04.8 & +05:26:30.99 & y $^{10}$ & 10.67 & $3.57_{-1.10}^{+1.34}\times10^{-14}$ \\
  J100645+462717 & 4.444 & 10:06:45.58 & +46:27:17.2 & c $^{25}$ &  & \\
  J101155-013052$^l$ & 5.58 & 10:11:55.54 & -01:30:52.69 & c $^{17}$ &  & \\
  J101335+281119 & 4.75 & 10:13:35.44 & +28:11:19.24 & y $^{10}$ &  & $<4.00\times10^{-14}$ \\
  J102107+220904 & 4.26 & 10:21:07.57 & +22:09:21.45 & y $^{8,16}$ & 9.26 & $2.81_{-0.96}^{+1.18}\times10^{-14}$ \\
  J102623+254259$^l$ & 5.25 & 10:26:23.62 & +25:42:59.44 & y $^{23}$ & 12.04 & $3.63_{-1.05}^{+1.26}\times10^{-14}$ \\
  J102838-084438 & 4.276 & 10:28:38.80 & -08:44:38.6 & y $^{33}$ & 85.55 & $2.46_{-0.28}^{+0.28}\times10^{-13}$ \\
  J103758+403328$^l$ & 6.07 & 10:37:58.18 & +40:33:28.74 & c $^{14}$ &  & \\
  J113350+481431$^l$ & 6.23 & 11:33:50.42 & +48:14:31.20 & c $^{14,1}$ &  & \\
  J114657+403708$^l$ & 5.005 & 11:46:57.79 & +40:37:08.66 & y $^{11}$ &  & \\
  J115503-310758 & 4.3 & 11:55:03.16 & -31:07:58.7 & y $^{20}$ & 27.74 & $4.21_{-0.84}^{+0.95}\times10^{-14}$ \\
  J123142+381658 & 4.137 & 12:31:42.17 & +38:16:58.9 & y $^{27}$ &  & \\
  J123503-000331 & 4.723 & 12:35:03.03 & -00:03:31.7 & y $^{27}$ &  & $<4.48\times10^{-14}$ \\
  J124413+862554$^l$ & 5.32 & 12:44:13.902 & +86:25:54.07 & c $^6$ &  & \\
  J125359-405930 & 4.464 & 12:53:59.53 & -40:59:30.7 & y $^{20}$ & 37.22 & $4.65_{-0.84}^{+0.95}\times10^{-14}$ \\
  J130738+150752 & 4.111 & 13:07:38.83 & +15:07:52.0 & c $^{25}$ &  & \\
  J130940+573309 & 4.28 & 13:09:40.70 & +57:33:09.9 & y $^{13}$ &  & \\
  J131121+222738 & 4.612 & 13:11:21.32 & +22:27:38.6 & c $^{25}$ &  & \\
  J132206-132354$^l$ & 4.71 & 13:22:06.46 & -13:23:54.71 & y $^7$ & 84.68 & $1.18_{-0.14}^{+0.14}\times10^{-13}$ \\
  J132512+112329 & 4.42 & 13:25:12.49 & +11:23:29.8 & y $^{13}$ & 17.98 & $3.01_{-0.83}^{+0.93}\times10^{-14}$ \\
  J134811+193520 & 4.4 & 13:48:11.25 & +19:35:23.64 & y $^{8,16}$ &  & \\
  J141212+062408 & 4.47 & 14:12:09.97 & +06:24:06.88 & y $^{8,16,27}$ & 32.50 & $4.46_{-0.84}^{+0.96}\times10^{-14}$ \\
  J142048+120545 & 4.02 & 14:20:48.01 & +12:05:45.9 & y $^{26}$ &  & \\
  J143023+420450$^l$ & 4.72 & 14:30:23.73 & +42:04:36.51 & y $^{32}$ &  & \\
  J143413+162852 & 4.195 & 14:34:13.05 & +16:28:52.7 & c $^{25}$ &  & \\
  J145459+110927 & 4.93 & 14:54:59.30 & +11:09:27.89 & y $^{10}$ &  & \\
  J151002+570243 & 4.31 & 15:10:02.92 & +57:02:43.4 & y $^{34}$ &  & \\
  J152028+183556 & 4.123 & 15:20:28.14 & +18:35:56.1 & c $^{25}$ &  & \\
  J153533+025419 & 4.39 & 15:35:33.88 & +02:54:23.38 & y $^{8,16}$ &  & \\
  J162956+095959 & 5.0 & 16:29:57.28 & +10:00:23.49 & y $^{8,16}$ &  & \\
  J164856+460341$^l$ & 5.36 & 16:48:54.53 & +46:03:27.3 & y $^{8,16}$ &  & \\
  J165913+210115$^l$ & 4.784 & 16:59:13.23 & +21:01:15.8 & y $^{27}$ &  & \\
  J170245+130104$^l$ & 5.466 & 17:02:45.31 & +13:01:02.28 & y $^{18,30}$ &  & \\
  J171103+383016 & 4.0 & 17:11:05.52 & +38:30:04.30 & y $^{8,9}$ &  & \\
  J171521+214547 & 4.01 & 17:15:21.26 & +21:45:31.65 & y $^{13}$ &  & \\
  J172007+602824 & 4.424 & 17:20:07.19 & +60:28:24.0 & c $^{25}$ &  & \\
  J172026+310431 & 4.67 & 17:20:26.68 & +31:04:31.6 & y $^{10,25}$ &  & \\
  J195135+013442 & 4.11 & 19:51:36.02 & +01:34:42.7 & y $^{15}$ &  & \\
  J213412-041909 & 4.334 & 21:34:12.02 & -04:19:09.7 & y $^{26}$ &  & \\
  J222032+002537 & 4.2 & 22:20:32.49 & +00:25:37.5 & y $^{26}$ &  & \\
  J231449+020146 & 4.11 & 23:14:48.71 & +02:01:51.08 & y $^{8,16,27}$ &  & \\
  J235758+140205 & 4.35 & 23:57:58.55 & +14:02:01.83 & y $^{8,16}$ &  & \\
\hline                                   
\end{tabular}
\end{table*}
%

\section{Data and analysis}
\label{Sec:data}

A complete characterization of blazars nature and emission profile needs a full electromagnetic spectrum coverage. 
We collected archival data from radio to high energies (where possible) for our whole blazar sample. 
Radio data are retrieved from several public radio catalogs, whose details are reported in Section \ref{Sec:data_radio}.
X-ray spectra from telescopes other than eROSITA are mainly retrieved from literature \citep{belladitta19,ighina19,ighina24,ighina25,sbarrato13a,sbarrato15,sbarrato22}, since typically they are obtained through dedicated observations and not all-sky catalogs. 
The rest of the multiwavelength coverage is collected from the data archive hosted at the Space Science Data Center (SSDC\footnote{\url{https://tools.ssdc.asi.it/SED/}}).

 \subsection{eROSITA}

We compute eROSITA upper flux limits \citep{tubin24} for the non-detected blazars using X-ray photometry on the eROSITA standard calibration data products, such as the counts image, background image, and exposure time. 
Values are reported in Table \ref{Table:sample}. 
We consider a circular aperture with a radius of $\sim30$\arcsec, which is given by a PSF encircled energy fraction of EEF = 0.75. We follow the Bayesian approach described by \cite{kraft91} and computed upper limits with a confidence interval of 99.87\% (corresponding to a one-sided $3\sigma$ level). To calculate energy conversion factors (ECFs), we assumed an absorbed power-law model with spectral index $\Gamma_X = 2$ and Galactic absorption $3\times10^{20}\rm \; cm^{-2}$. The upper limits are provided in the most sensitive eROSITA band (0.2--2.3 keV) of the deepest eROSITA stacked observations (i.e., eRASS:4 and eRASS:5 when available).

We analyze the eROSITA spectral properties for the 8 brightest detected blazars. We extract the spectra from the pipeline-processed eventfiles \citep[version 020,][]{brunner02}, combining events from all seven telescope modules and from all completed eROSITA All Sky Surveys (eRASS:5). We use the \texttt{srctool} task of the eROSITA Science Analysis Software System (eSASS) to extract source and background spectra, ancillary response files (ARFs), and response matrix files (RMFs). Source counts are extracted from circular regions with radii of 60", and background counts from large, nearby, and source-free regions. 

In the case of J052506-334305, J032444-291821, J102838-084438, and J132206-132354, where we have $>$80 total counts per source, we also extract and analyze the spectra of each individual eRASS. Given that these sources are bright, we use the pipeline-extracted source and background spectra, ARFs, and RMFs \citep[see][for a detailed description of the pipeline extraction]{tengliu22}. The eRASS3 observations of J132206-132354 and the eRASS4 observations of J032444-291821 yielded no detection; therefore, we computed single-epoch upper limits with the same prescriptions as for the undetected sources.

All spectra were analysed with the Bayesian X-ray Analysis software (BXA) version 4.1.2 \citep{buchner2014bxa}, which connects the nested sampling algorithm UltraNest \citep{buchner2019ultranest,buchner2021ultranest} with the fitting environment CIAO/Sherpa \citep{fruscione2006sherpa}. Spectra were fit unbinned and using Cash-statistics. Our fitting approach simultaneously models source and background spectra. For the background, we used a principal component analysis-based background model \citep{simmonds2018} derived from a large sample of eROSITA background spectra. We instead modeled all source spectra as powerlaws absorbed by Galactic absorption foreground, for which we estimate the contribution through the results of the HI4PI collaboration \citep{Hi4pi}. The results of the analysis for the cumulative eROSITA observations are reported in Table \ref{Table:ero-spec-fit}, while the time-resolved spectral properties are shown in Fig. \ref{Fig:variability}.

%
\begin{table}
\renewcommand{\arraystretch}{1.5} 
\caption{X-ray spectral analysis for the 8 brightest eROSITA sources.}            
\label{Table:ero-spec-fit}      
\centering                          
\begin{tabular}{l c l}        
\hline\hline 
 Name & $\Gamma_X$ & $F_{[0.2-2.3 {\rm keV}]}$ \\ 
     &          & erg/cm$^2$/s  \\ 
\hline                        
J032444-291821 & $1.74^{+0.26}_{-0.27}$ & $1.18_{-0.17}^{+0.19}\times10^{-13}$  \\
J052506-334305 & $1.46^{+0.11}_{-0.11}$ & $4.87_{-0.33}^{+0.34}\times10^{-13}$  \\
J083548+182519 & $1.54^{+0.46}_{-0.43}$ & $1.24_{-0.30}^{+0.35}\times10^{-13}$  \\
J085111+142337 & $1.14^{+0.45}_{-0.47}$ & $1.12_{-0.27}^{+0.31}\times10^{-13}$  \\
J102838-084438 & $1.62^{+0.27}_{-0.27}$ & $2.87_{-0.42}^{+0.48}\times10^{-13}$  \\
J125359-405930 & $2.19^{+0.55}_{-0.52}$ & $7.48_{-2.29}^{+2.97}\times10^{-14}$  \\
J132206-132354 & $1.12^{+0.27}_{-0.28}$ & $1.23_{-0.18}^{+0.20}\times10^{-13}$  \\
J141212+062408 & $2.07^{+0.68}_{-0.74}$ & $3.71_{-1.73}^{+2.07}\times10^{-14}$  \\
\hline                                   
\end{tabular}
\end{table}
%

%
\begin{table}
\renewcommand{\arraystretch}{1.3} 
\caption{Spectral analysis of each eRASS scan of the 4 eROSITA sources with more than 80 counts. }            
\label{Table:X-spectra}      
\centering  
\small
\begin{tabular}{l l c l}        
\hline\hline 
 & scan &$\Gamma_X$ & $F_{[0.2-2.3 {\rm keV}]}$ \\ 
     &    &     & erg/cm$^2$/s  \\ 
\hline                        
\multicolumn{3}{l}{\it J032444-291821} & \\
  &  eRASS1    & $1.82^{+0.38}_{-0.39}$    & $1.20_{-0.26}^{+0.31}\times10^{-13}$ \\
  &  eRASS2    & $2.15^{+0.48}_{-0.44}$    & $9.26_{-2.53}^{+3.07}\times10^{-14}$ \\
  &  eRASS3    & $1.26^{+0.51}_{-0.52}$    & $8.38_{-2.49}^{+3.04}\times10^{-14}$ \\
  &  eRASS4    & \multicolumn{2}{c}{\tiny undetected}                          \\  
  &  eRASS5    & $1.76^{+0.47}_{-0.47}$    & $1.23_{-3.09}^{+3.56}\times10^{-13}$ \\
\hline                                   
\multicolumn{3}{l}{\it J052506-334305} & \\
  &  eRASS1    & $1.00^{+0.21}_{-0.20}$    & $3.79_{-0.42}^{+0.48}\times10^{-13}$ \\
  &  eRASS2    & $1.53 \pm 0.18$           & $4.30_{-0.49}^{+0.50}\times10^{-13}$ \\
  &  eRASS3    & $1.36 \pm 0.18$           & $5.09_{-0.57}^{+0.60}\times10^{-13}$ \\
  &  eRASS4    & $1.65^{+0.18}_{-0.17}$    & $5.33_{-0.55}^{+0.59}\times10^{-13}$ \\
  &  eRASS5    & \multicolumn{2}{c}{\tiny not scanned}                          \\
\hline                                   
\multicolumn{3}{l}{\it J102838-084438} & \\
  &  eRASS1    & $2.56^{+0.46}_{-0.43}$    & $4.30_{-1.09}^{+1.41}\times10^{-13}$ \\
  &  eRASS2    & $2.15^{+0.46}_{-0.43}$    & $3.74_{-0.88}^{+1.13}\times10^{-13}$ \\
  &  eRASS3    & $1.37 \pm 0.41$           & $3.38_{-0.71}^{+0.84}\times10^{-13}$ \\
  &  eRASS4    & $1.34 \pm 0.51$           & $1.80_{-0.51}^{+0.65}\times10^{-13}$ \\
  &  eRASS5    & \multicolumn{2}{c}{\tiny not scanned}                          \\
\hline                                   
\multicolumn{3}{l}{\it J132206-132354} & \\
  &  eRASS1    & $0.94^{+0.55}_{-0.50}$    & $1.33_{-0.37}^{+0.44}\times10^{-13}$ \\
  &  eRASS2    & $1.40 \pm 0.37$           & $1.83_{-0.38}^{+0.51}\times10^{-13}$ \\
  &  eRASS3    & \multicolumn{2}{c}{\tiny undetected}                          \\ 
  &  eRASS4    & $1.40^{+0.60}_{-0.57}$    & $1.04_{-0.34}^{+0.43}\times10^{-13}$ \\
  &  eRASS5    & $0.10^{+0.55}_{-0.50}$    & $9.32_{-2.74}^{+3.66}\times10^{-14}$ \\
\hline                                   
\end{tabular}
\end{table}
%

 \subsection{Radio}
 \label{Sec:data_radio}
We cross-matched our catalog of blazars with several public radio surveys.
In the specific we used: 
the LOw-Frequency ARray (LOFAR) Two-metre Sky Survey (LoTSS) second data release (\citealt{shimwell2022}) at 144~MHz, 
the TIFR GMRT Sky Survey (TGSS, \citealt{intema2017}) at 150~MHz,
the Sydney University Molonglo Sky Survey (SUMSS; \citealt{bock1999,mauch2003}) at 843~MHz,
the Rapid ASKAP Continuum Survey (RACS, \citealt{mcconnell2020,hale2021,duchesne2023}) at 888~MHz, 1.367 and 1.65~GHz,
the Evolutionary Map of the Universe survey (EMU, \citealt{norris2011}) at 943~MHz,
the Faint Images of the Radio Sky at Twenty-cm (FIRST, \citealt{becker1994}), the NRAO VLA Sky Survey \citep[NVSS,][]{condon1998} and 
the MeerKAT Absorption Line Survey (MALS, \citealt{gupta2016}) all at 1.4~GHz and 
the VLA Sky Survey (VLASS, \citealt{lacy2020,gordon2021}) at 3~GHz.
Given that many of these sources are well-known and studied in the literature, data reported in dedicated follow-up papers have also been used, which may include, for example, data from the update Giant Metrewave Radio Telescope (uGMRT) and the Very Large Array (VLA) at different frequencies.  

Additionally, we examined the radio images to exclude any flux densities affected by known issues, such as the well-documented phase errors present in VLASS data\footnote{\url{https://science.nrao.edu/vlass/data-access/vlass-epoch-1-quick-look-users-guide}}.
In case the sources appear extended we have re-computed the flux densities by performing a Gaussian fit with the task \texttt{IMFIT} of the Common Astronomy Software Applications package (CASA, \citealt{mcmullin2007}). 
These particular cases are reported in Sect. \ref{sec:notes}. 

\section{Characterization}
\label{Sec:character}

 \subsection{eROSITA variability}

\begin{figure}
   \centering
   \includegraphics[width=1.05\hsize]{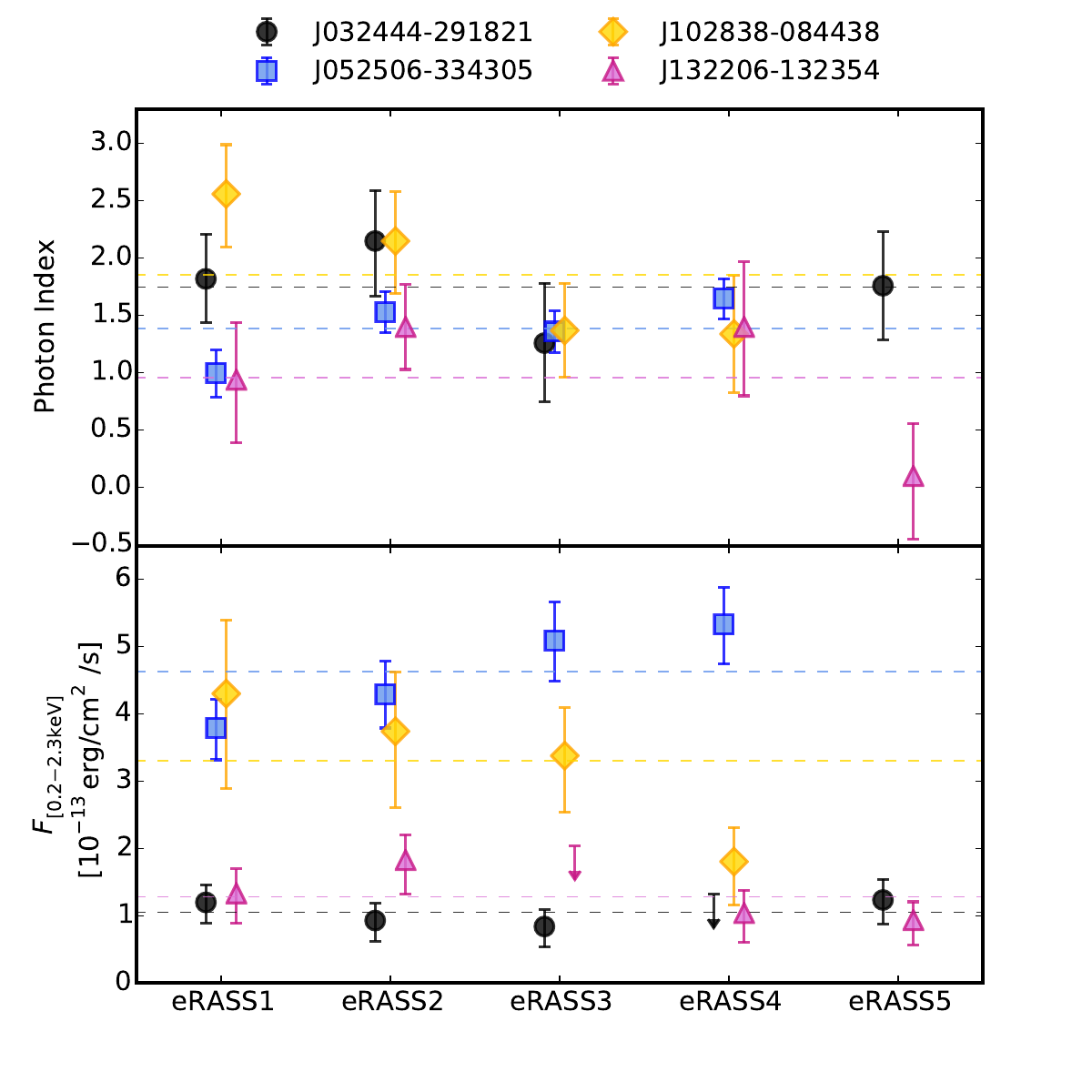}
      \caption{eROSITA photon index (top panel) and X-ray flux 
               in the 0.2-2.3 keV range (bottom panel)
               of the 4 sources with more than 80 counts, 
               as a function of the eRASS scan 
               (i.e.\ roughly function of time), as labelled. 
               The horizontal dashed lines show the mean value of each source (photon index or flux, in the respective panels) as color-coded, calculated excluding upper limits.
              }
         \label{Fig:variability}
   \end{figure}

J052506-334305, J032444-291821, J102838-084438, and J132206-132354 have more than 80 counts per source in eROSITA data. 
It was thus possible to extract spectra for each individual eRASS (Table \ref{Table:X-spectra}).
Figure \ref{Fig:variability} shows the temporal evolution of photon index and observed flux of the four sources. 
Both parameters do not show any significant variability for any of the source. 
J102838-084438 only shows a flux variation at one sigma level between eRASS3 and eRASS4. 
No confirmation of its variability could be done, since the source was not covered by eRASS5. 

Checking for blazar variability, is a powerful tool provided by the eROSITA multiple sky scanning. 
Associated to the instrument's sensitivity, this search is allowed even at high redshift, as demonstrated by this small set of bright sources. 

%

 \subsection{Broad-band SED}
 \label{sec:SED}

A multi-wavelength SED, compiled across the whole electromagnetic spectrum, is crucial to interpret blazar physics and geometry. 
Because of its strong relativistic beaming, jet emission significantly changes not only depending on its intrinsic features, but also on its orientation with respect to the observers line of sight. 
A plethora of models based mainly on different particle compositions have been developed in the last 20-30 years, from purely leptonic to lepto-hadronic models, single or multiple emitting regions \citep[see e.g.][for a review]{boettcher19}. 
The debate on their validity being still open, a good multi-wavelength coverage is needed to discriminate among them, or at least put them to the test. 
At $z>4$, the data coverage is generally not enough to do that, mainly because the information provided by $\gamma$-ray observatories on the Inverse Compton power and peak emission are missing. 
Without reliable observational constraints, there is no point in performing a numerical, physical-based modeling, and therefore in the following we consider the parametric, phenomenological SED developed by \cite{ghisellini17}. 
The authors developed an analytic approximation of the $\gamma$-detected blazars, composed by a series of broken power-laws. 
They extracted them from the SEDs of all {\it Fermi}/LAT-detected blazars, averaged on different $\gamma$-ray luminosity bins. 

We derive ranges on key observational parameters that are tightly connected to the jet orientation and its overall power, i.e.\ the Compton dominance, synchrotron peak position, and power-law spectral indices that trace the IC hump intensity in the SED. 
In the case of the 10 candidates, no X-ray data are available, and therefore no constrain on the IC features can be obtained. For this reason, no SED modeling is performed on the 10 candidates.

The current high-$z$ blazar selection criteria require that all sources have rest frame optical-UV features consistent with quasars, i.e.\ their emission in this band is not jet-dominated but most likely thermal, and they show strong broad emission lines to provide for redshift estimates. 
We have thus a chance to model their big blue bumps with accretion disk emission models, in order to derive rough estimates on their supermassive black hole masses and accretion rates. 
A simple, analytic \cite{shakura73} model is here implemented, along with an anisotropy correction due to the likely disc geometry \citep{calderone13}.

The sparse data coverage does not allow for tight constraints on the model parameters. 
We thus derive parameter ranges whose limits are extracted from a lower and a higher flux solution encompassing all the SED profiles possibly consistent with the available data. 
The ranges of values are reported in the Appendix, in Tables \ref{Table:SED-results1} and \ref{Table:SED-results2}.
All SED models are consistent with powerful FSRQs profiles, more extreme in terms of synchrotron peak position and Compton Dominance than the reddest SEDs found by \cite{ghisellini17}. 
This is a trend already clearly highlighted by modelling $z>4$ blazars with physical-based, numerical SEDs \citep[see e.g.\ the fit to the broad-band spectrum of archetypical blazar B2~1023+25][]{sbarrato12,sbarrato13b}, and our results confirm it.

\subsection{Radio spectrum and Radio loudness}
\label{Sec:radiospecR}
All the blazars discussed in this paper are previously known sources that have been studied in the literature—some for several decades. For example, the radio properties of many of these objects have been investigated since the 1990s, when the first all-sky radio surveys became available. In some cases, detailed studies and targeted follow-up observations have been conducted.

However, the wealth and diversity of existing data have made it challenging to compile consistent information on their radio properties. To address this, we adopted the following approach to determine the radio spectral indices—and, consequently, the radio loudness and radio luminosity—of all blazars and candidates in our sample.

For sources that have been the subject of detailed follow-up studies (particularly those using simultaneous or recent observations), we adopted the spectral indices reported in the literature. These are marked with an "$l$" in the table \ref{Table:sample}. 
For the remaining sources, we calculated the spectral index by combining radio data from both recent and archival surveys (see Sect. \ref{Sec:data_radio} for details).

Each radio SED was modeled using either a single power-law or a broken power-law, depending on its shape (e.g., flat, steep, inverted, peaked, or complex spectra). We note that most sources do not have simultaneous radio flux measurements, and variability may affect the spectral index estimates. 
Whenever a flux measurement was suspected to be influenced by variability, it was excluded in favor of data points that were closer in time. A detailed analysis of radio variability is beyond the scope of this paper.

The radio SED analysis reveals diverse spectral properties across the sample. 
Of the 62 sources, 44 can be fitted with a single power-law, while 18 require a double power-law model. Among the latter, 12 exhibit a peaked spectrum, 2 show a convex spectrum, and 4 display a combination of flat and steep components. We do not currently observe any correlation between these different spectral behaviors and radio luminosity, radio loudness, redshift, or mass, nor has such a trend been reported in the literature .

The radio loudness $R$ is a parameter that quantifies how powerful is the synchrotron radiation produced by the jet with respect to the optical emission associated with the accretion disk. 
In this paper we used the definition of \cite{kellermann89}: $
    R=f_{5 \rm ~GHz}/f_{4400\AA}$
which is widely used in the literature. 

Blazars usually show very high values ($>100$) of R, for example, \citealt{sbarrato13a,belladitta19,belladitta20,caccianiga19}, but they cover a wide range that also depends on which radio and optical values are used to estimate this parameter. 
To obtain consistent results for the entire sample, we decided to recalculate the $R$ values as uniformly as possible. 
The 5~GHz rest frame flux density correspond to an observed frequency between $\sim$0.6 and 1~GHz for sources in the redshift range 4-7. Therefore it was estimated starting from the flux density observed in RACS (888~MHz) or FIRST (1.4~GHz) and using the radio spectral index previously calculated.
 
The 4440$\AA$ rest frame flux density was computed from the $z$-band magnitude and using the optical spectral index of \cite{vandenberk01} ($\alpha_{\nu}$ = 0.44).
Magnitudes are from the Panoramic Survey Telescope and Rapid Response System (Pan-STARRS, PS1, \citealt{chambers2016}), the Dark Energy Survey (DES, \citealt{flaugher2005,abbott2018}) and the DESI Legacy Survey (DELS; \citealt{dey2019}). 
However, if the optical magnitudes from PS1, DES, and DELS were very different from each other, the R was calculated as the average of the values estimated using the individual magnitudes.
Differences in optical magnitude can be a sign of source variability.

   \begin{figure}
   \centering
   \includegraphics[width=1.05\hsize]{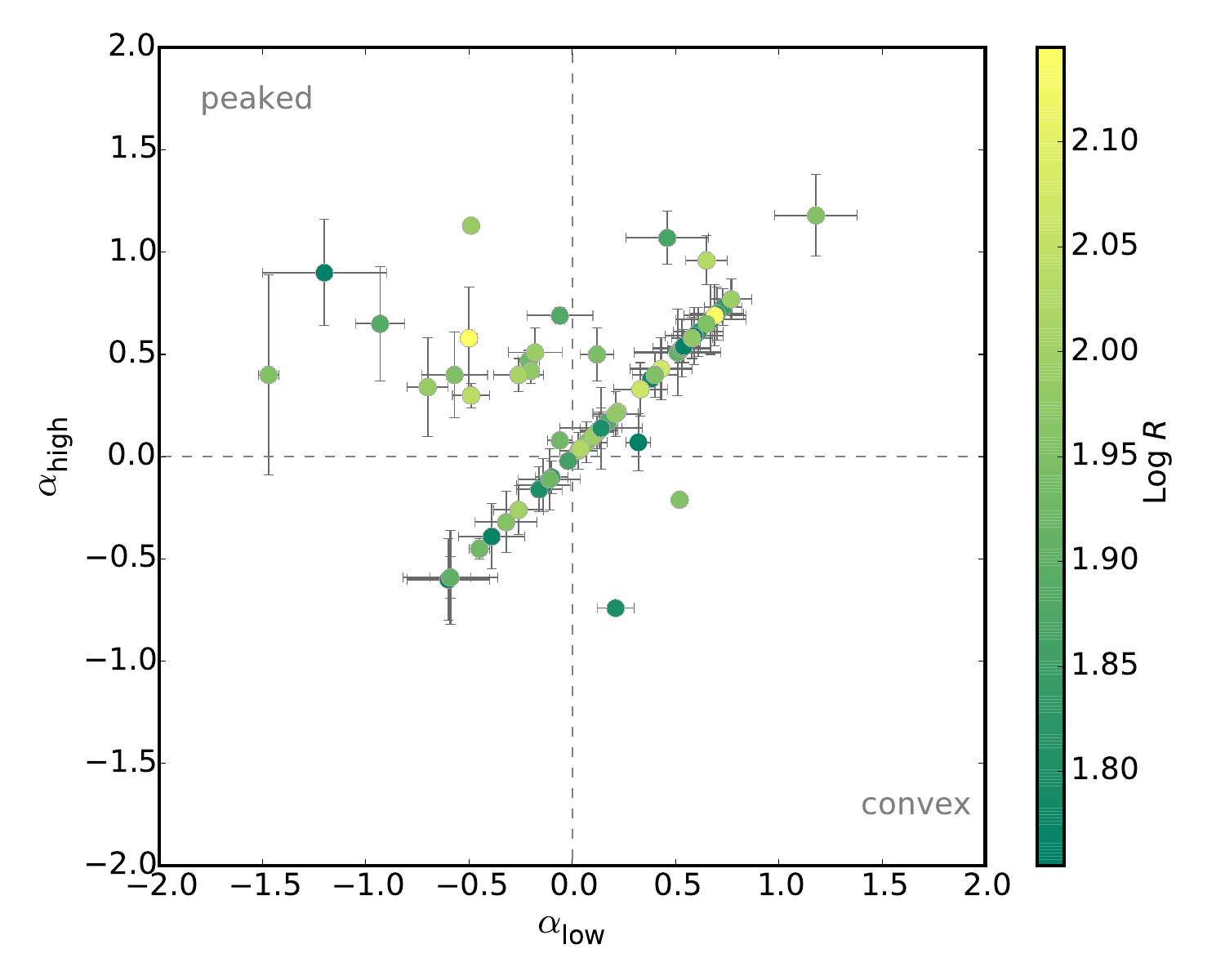}
      \caption{Radio spectral index distribution of the extended sample, including both confirmed blazars and candidates, color coded for their radio loudness $R$. 
      The data distributed along the identity diagonal show results for sources best fitted with a single power-law, while those best described by double power-laws are located in the first or last quadrants depending on whether they have peaked or convex radio spectra, as labeled.
      }
      \label{Fig:radio_spectra}
   \end{figure}
%

\section{Notes on sources}
In this section we report some particular cases we encountered while performing the radio and/or X-ray analysis of the blazars in our sample. 
The sources are listed in order of R.A.
\label{sec:notes}

\begin{itemize}
    \item[-] \textbf{J001115$+$144601}: The FIRST 1.4~GHz, RACS 1.65~GHz and VLASS 3~GHz radio images clearly show the presence of three distinct radio sources: the blazar, a component to the north (N) and one to the east (E) at 16.3 and 29.5~arcsec apart, respectively.  
    Therefore, the radio flux densities from lower resolution surveys/images, such as TGSS (25$''$), RACS 888~MHz (25$''$) and NVSS (45$''$), are contaminated by the presence of these additional objects. To estimate the radio spectrum of J001115$+$144601, these three radio data points were removed, because no reliable decomposition of the closest source could be done. 
    The N and E objects are not coincident with any optical/infrared counterpart and have no X-ray association. 
    The radio spectral index of both the N and E components are steep ($\alpha_{r\_E}=0.8$ and $\alpha_{r\_N}=1.2$, which is characteristic of radio lobes (e.g., \citealt{deGasperin2018}). Such features are typically associated with remnant emission from past AGN activity (e.g., \citealt{shulevski2012}). 
    However, the steep spectrum is also consistent with a high-redshift radio galaxy (HzRG, \citealt{debreuck2010}), which could remain obscured in optical bands but detectable in deep infrared or sub-millimeter observations (e.g., \citealt{reuland2003}).     
    With this data in hand, we are not able to rule out the nature of these two sources,  whether one of the two components is an extended emission associated with the blazar or whether they are both independent objects.
            
    \item[-] \textbf{J021043$-$001818}: At the resolution of the NVSS and RACS 888~MHz, this source appears compact and unresolved. However, the higher resolution radio images from FIRST ($\sim$5$''$) and VLASS ($\sim$2.5$''$) reveal a close extension in the NE direction (detected in all the VLASS epochs). This is consistent with what found by \cite{coppejans16} that reports a resolved radio emission at 5 GHz observed at $\sim$1.5$''$ of resolution with the European VLBI Network (EVN). No optical/infrared counterparts at the position of the extended radio emission have been found. In our radio analysis we have used only the flux densities of the central core, estimated with CASA.

    \item[-] \textbf{J030437$+$004653}: This source is classified as a blazar based on high-resolution radio data (\citealt{krezinger2022}): it has a flat radio spectrum with Doppler boosted emission, and shows a potential signal of variability.
    However, the lack of detection in {\it Swift}/XRT suggests that the jet is not pointed towards us in X-rays (\citealt{sbarrato22}).
    \cite{zhu19} observed the source with {\it Chandra} and reported a faint detection, with no possibility of deriving a strong constrain on the X-ray photon index. Their result, derived with an upper limit on the photon index ($>1.79$), is well consistent with the {\it Swift}/XRT upper limit, and despite being 1.5 order of magnitudes brighter than the expected X-ray Corona emission for a $\sim2.5\times10^{46}$erg/cm$^2$/s, does not describe the standard, hard X-ray spectrum typical of a high-$z$ blazar IC component. 
    It might instead be emitted by a region accelerated along a slightly misaligned direction, and thus slightly de-beamed with respect to a bright X-ray blazar.
    If this is the case, this source might be the differently aligned counterpart of the X-ray blazars with misaligned radio components (see below J222032$+$002537). 
    The emission traced by NVSS is however contaminated by two nearby radio sources that become clearly separated in the higher resolution RACS 1.65~GHz image. Based on the DESI database\footnote{\url{https://www.legacysurvey.org/viewer}}, these two objects are potentially associated with a $z\sim0.9$ galaxy and a $z\sim2.4$ quasar.

    \item[-] \textbf{J032214$-$184117}: No detection of this source is present in eRASS:5, but eRASS2 (i.e.\ the data collected during the second complete eROSITA sky coverage) shows the presence of an X-ray source at a distance of 15 arcsec from its optical position, with $F_{\rm [0.2-2.3keV]}=2.05^{+1.26}_{-0.97}\times10^{-14}{\rm erg/cm^2/s}$ from $\sim4$ counts.  
    This might suggest that during eRASS2 the source produced an X-ray flare, or the background signal in the cumulative data does not allow for a secure source detection. \cite{ighina25} observed the source with {\it Chandra}, obtaining a flux $F_{\rm [0.5-7keV]}=1.07^{+0.13}_{-0.04}\times10^{-14}{\rm erg/cm^2/s}$ with $\Gamma=2.1\pm0.5$. 

    \item[-] \textbf{J052506$-$334305}: Older measures of its radio flux densities (SUMSS and NVSS) show variability with respect to the more recent RACS. 
    This source had already been detected in X-rays \citep[][]{worsley04}, and classified as blazar for its hard X-ray spectrum and strong, flat radio spectrum. 
    The eROSITA observations result in an even harder X-ray spectrum ($\Gamma_X=1.46\pm0.11$) that might suggest variability in this frequency range as well. 
    As already explained in Section \ref{sec:SED}, the data coverage does not allow for a statistically significant fit of the broad-band SEDs, and hence we derive ranges of validity for the phenomenological model we use. 
    In the case of this source, it must be noted that the parameters reported in Table \ref{Table:SED-results1} describe a much softer IC bump than what is described by eROSITA data. 
     \item[-] \textbf{J081333$+$350810}: This object exhibits extended radio emission in the north-west direction, detected in LOFAR 144~MHz, FIRST, and all VLASS epochs. The separation between the two components is 6.8~arcsec.
    The extended component is brighter in LOFAR (117~mJy) than the central core (105~mJy), but its flux density decreases significantly (by a factor of $\sim$10) at higher frequencies.
    Its steep radio spectral index ($\alpha_r=1.13$) is typical of radio lobes.
    This interpretation is supported by the absence of optical/IR counterparts at the position of the extended emission in wide-field surveys such as PanSTARRS and DELS. However, the steep spectrum is also consistent with a HzRG. With the available data, we cannot definitively distinguish between these scenarios.
     \item[-] \textbf{J123142.17$+$381658.9}: The FIRST radio image reveals two possible extensions, one to the south and one to the south-west. However, the image angular resolution (5$''$) prevents clear separation of these potential components from the core. These extensions appear to be confirmed by the higher-resolution VLASS 3.1 image (2.5$''$), where they are more clearly resolved. If both extensions are associated with the source, this morphology would be consistent with a \textit{wide-angle tail radio galaxy} (WAT, \citealt{owen1976}). WATs are characterized by initially well-collimated jets on kiloparsec scales that subsequently flare into diffuse plumes, which may exhibit significant bending due to the motion of the jet through the intergalactic medium \citep[e.g.,][]{owen1976}. The absence of optical/NIR counterparts at the positions of the two extensions further supports their association with the radio source. 
     Higher resolution radio observations (e.g., VLBI) are needed to unveil the real nature of these extended components. 
    \item[-] \textbf{J130738$+$150752}:  This is similar to the case of J081333$+$350810. At the resolution of TGSS, RACS 888~MHz, and NVSS, the source appears compact and unresolved. However, a second component in the north-east (NE) direction emerges in the higher resolution radio images from FIRST, RACS 1.65~GHz, and VLASS. There is a hint of extension in the RACS 1.37 GHz image, but without a clearly separated source. 
    The NE component is significantly brighter than the main source: $\sim$3 times brighter in FIRST (12.5~mJy vs. 3.6~mJy) and RACS 1.65~GHz (9.6~mJy vs. 3~mJy), and $\sim$2 times brighter in VLASS (4~mJy vs. 2~mJy). No optical counterpart is detected at the position of the NE component in optical surveys. Therefore, we cannot determine whether this is an independent background/foreground object or extended radio emission associated with the blazar.
    \item[-] \textbf{J170245$+$130104}: this source is the first blazar discovered in the eROSITA Russian part of the Sky (\citealt{khorunzhev21}). \cite{tao23} identified a radio component in the south-west (SW) as a foreground galaxy at z$\sim$0.677, based on photometric redshift estimates using SDSS and WISE data. The radio flux of J1702$-$SW contaminates the flux of the blazar at low-frequency and at low-resolution radio images (e.g., NVSS).
    \item[-] \textbf{J222032$+$002537}: \cite{cao17} suggested that this source is not enhanced by the Doppler effect, i.e. it is not a blazar, based on EVN radio data at 1.4 GHz. The extended emission on arcsec scale and Doppler factor well below 1 could indicate a double-lobed radio AGN with a larger inclination relative to the line of sight. The extended emission is also visible in FIRST and VLASS. The optical position of the blazar does not overlap with the brightest part of the radio emission in these two images, indicating that the optical AGN is located at the center of the asymmetric radio emission, supporting \cite{cao17} hypothesis.  
    However, X-ray properties reported in the literature \citep{sbarrato15} show, in contrast, clear blazar features. 
    The reason of this stark discrepancy has already been thoroughly debated \citep[e.g.\ ][]{sbarrato15,sbarrato22,cao17,caccianiga19}, together with slight discrepancies in the apparent orientation shown by other sources \citep{cao17}.
    The main viable options include (i) a possible bending of the jet during its propagation, at first pointing toward us and then being misaligned up to a full radiogalaxy orientation,  (ii) re-orientation of the jet from misaligned to aligned with respect to our line-of-sight, due to reorientation of the central engine or nuclear system, or (iii) intrinsic differences in acceleration, emission and/or propagation of the jet at very high redshift. 
    Even if small differences are present (e.g.\ more frequent compact or inverted radio spectra), they are usually ascribed to the young jet ages, and the overall jetted AGN sample at $z>4$ does not seem to suggest drastic differences. 
    A re-orientation of the jet base due to intrinsic changes in the SMBH and accretion system would require longer timescales (that are not allowed easily at such high $z$), but cannot be excluded. 
    Bending of the jet is a viable option, but the details are not yet clear.    
\end{itemize}

\section{Population}
\label{Sec:population}

 \subsection{Space Densities}

%
\begin{table*}
\caption{Comoving number densities ($\Phi(z)$) derived for all BLAZ4R sources, only those with $M>10^9M_\odot$ or $L_{\rm bol}>10^{47}$erg/s, and the equivalent sub-samples with eRASS:5 detections. All values are derived for the redshift bins as detailed in each column, and for three different bulk Lorentz factor values: 13, 10, 8.}            
\label{Table:Phi}      
\renewcommand{\arraystretch}{1.1} 
\centering                          
\begin{tabular}{l c c c c c c}        
\hline\hline 
    & $\Gamma$ & $4\leq z<4.7$ & $4.7\leq z<5.4$ & $5.4\leq z<6.1$ & $6.1\leq z<6.8$ & $6.8\leq z<7.5$ \\ 
       &   & [Gpc$^{-3}$] & [Gpc$^{-3}$] & [Gpc$^{-3}$] & [Gpc$^{-3}$] & [Gpc$^{-3}$] \\ 
\hline                        
 BLAZ4R & 13 & 33.6 & 19.3 & 7.8 & 1.4 & 1.5 \\      
             & 10 & 19.9 & 11.4 & 4.6 & 0.8 & 0.9 \\      
             & 8  & 12.7 & 7.3 & 3.0 & 0.5 & 0.6 \\      
 \vspace{0.05cm}
             & 4  & 3.2 & 1.8 & 0.7 & 0.1 & 0.1 \\      
 $\;M>10^9M_\odot$ & 13 & 28.0 & 14.5 & 5.2 & 1.4 & - \\      
             & 10 & 16.6 & 8.6 & 3.1 & 0.8 & - \\      
             & 8  & 10.6 & 5.5 & 2.0 & 0.5 & - \\      
 \vspace{0.05cm}
             & 4  & 2.7 & 1.4 & 0.5 & 0.1 & - \\      
 $\;L_{\rm bol}>10^{47}$erg/s & 13 & 25.8 & 12.1 & 3.9 & - & 1.5 \\      
             & 10 & 15.2 & 7.2 & 2.3 & - & 0.9 \\      
             & 8  & 9.8 & 4.6 & 1.5 & - & 0.6 \\      
 \vspace{0.1cm}
             & 4  & 2.4 & 1.1 & 0.4 & - & 0.1 \\      
 \hline   
 eRASS:5 detected & 13 & 26.9 & 12.1 & 5.2 & - & - \\
                  & 10 & 15.9 & 7.2 & 3.1 & - & - \\
                  & 8  & 10.2 & 4.6 & 2.0 & - & - \\      
 \vspace{0.05cm}
                  & 4  & 2.5 & 1.1 & 0.5 & - & - \\      
 $\;M>10^9M_\odot$ & 13 & 24.6 & 9.7 & 2.6 & - & - \\      
             & 10 & 14.6 & 5.7 & 1.5 & - & - \\      
             & 8  & 9.3 & 3.7 & 1.0 & - & - \\      
 \vspace{0.05cm}
             & 4  & 2.3 & 0.9 & 0.2 & - & - \\      
 $\;L_{\rm bol}>10^{47}$erg/s & 13 & 20.2 & 4.8 & 2.6 & - & - \\      
             & 10 & 11.9 & 2.9 & 1.53 & - & - \\      
             & 8  & 7.6 & 1.8 & 1.0 & - & - \\      
 \vspace{0.1cm}
             & 4  & 1.9 & 0.5 & 0.2 & - & - \\      
\hline                                   
\end{tabular}
\end{table*}

   \begin{figure}
   \centering
   \includegraphics[width=\hsize]{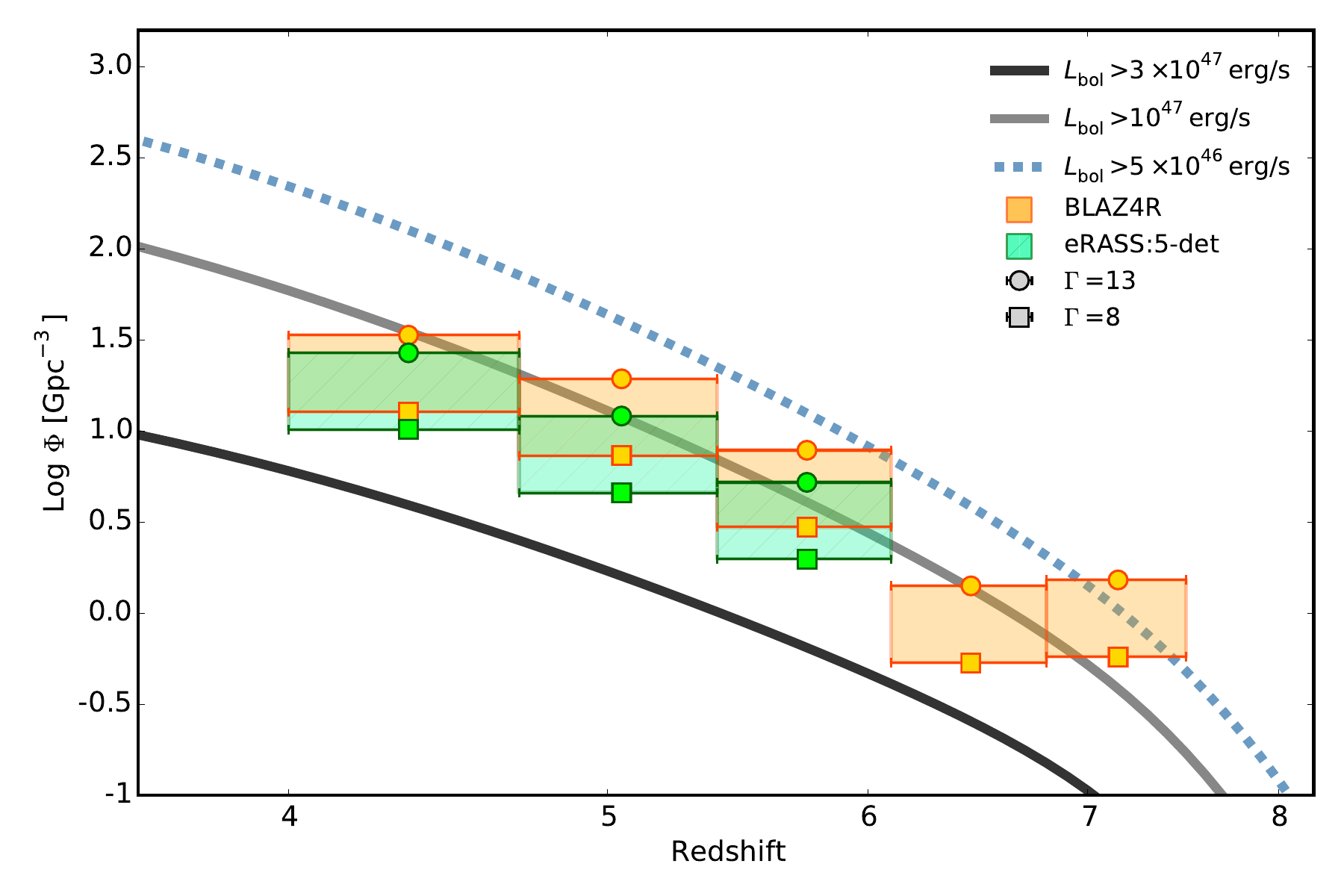}
   \includegraphics[width=\hsize]{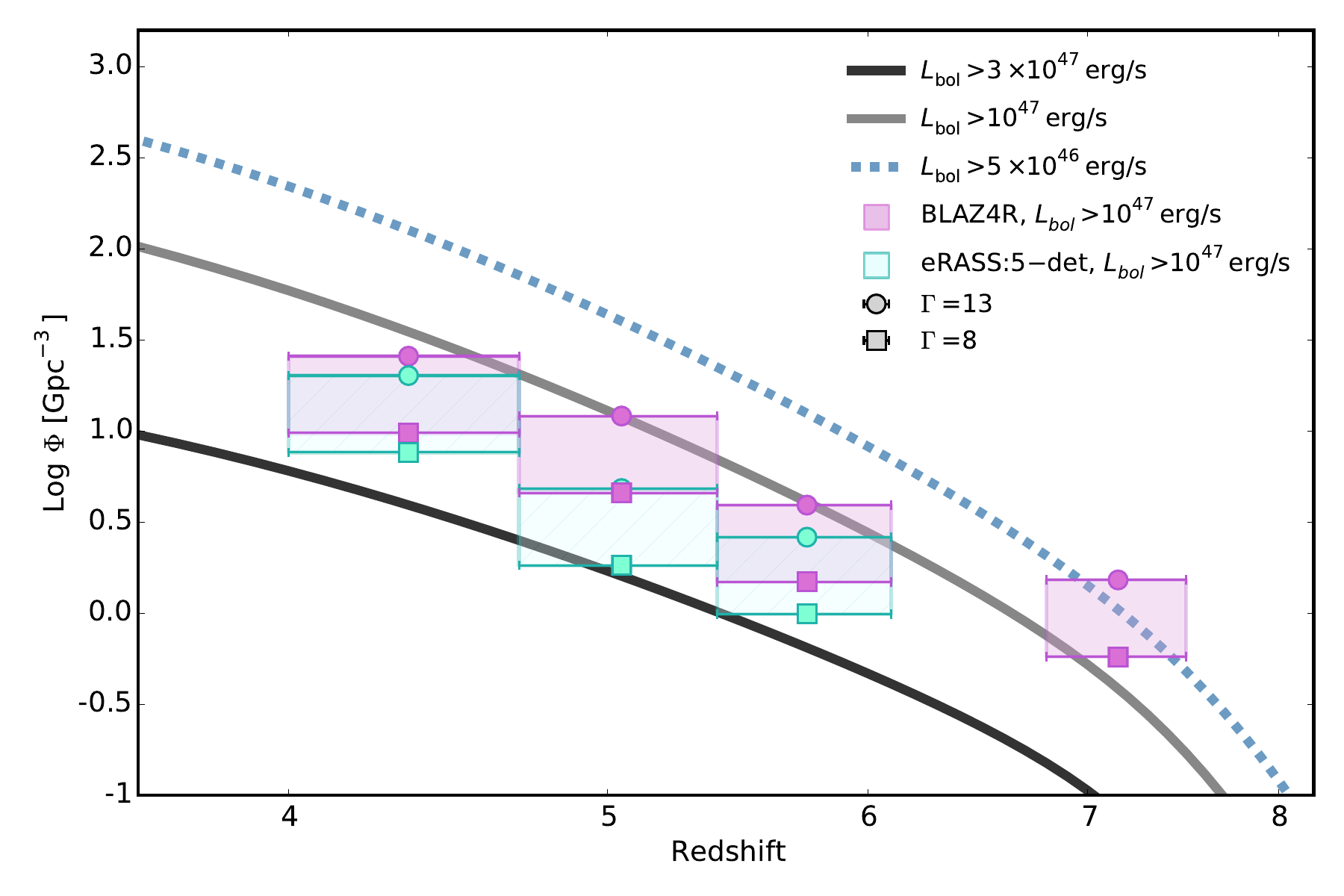}
      \caption{Comoving number densities of active black holes 
               hosted in jetted AGN traced by blazars at $z>4$, 
               as a function of redshift. 
               $\Phi(z)$ is derived for the jetted population by assuming 
               a bulk Lorentz factor of $\Gamma=13$ or 8 
               (circles or squares respectively).
               The upper figure shows the results for the whole sample, while the lower is limited to $L_{\rm bol}>10^{47}$erg/s. 
               In both figures, differently colored data points and shaded intervals refers to the total $z>4$ sample (orange in the top figure, purple in the bottom one), and the eRASS:5 detected only (green and light blue).
               The curves show the space densities for the overall 
               massive AGN population as extracted from \cite{shen20}, 
               integrated starting from a bolometric luminosity of 
               $5\times10^{46}, 10^{47}$ and $3\times10^{47}$ erg/s 
               (dashed blue, solid grey and solid black lines).
              }
         \label{Fig:Phi-zoom}
   \end{figure}
%

    Blazar peculiar orientation allows to use them as statistically significant tracers of the overall jetted AGN population.
    By defining them as those jetted AGN observed at viewing angles smaller than their jet beaming angles ($\theta_{\rm v}\leq\theta_{\rm b}=1/\Gamma$), the presence of $2\Gamma^2$ analogous jetted AGN with same intrinsic properties and different orientation can be inferred by each blazar detection \citep[][]{ghisellini10}. 
    Lorentz factors ranging between 8 and 15 ensure that for each blazar observed, $\sim130-450$ analogous accreting SMBHs hosted in jetted sources exist in the same redshift range. 
    We applied this so-called ``blazar argument" to our sample, in order to derive the space density of active SMBHs in jetted AGN and compare it with the statistics obtained from quasar luminosity functions \citep[][]{shen20}.

    Table \ref{Table:Phi} details the comoving number densities of the $z>4$ jetted AGN population in 5 redshift bins and for different $\Gamma$ values. 
    The data coverage of the sources in our sample does not allow for a numeric, physically consistent SED fitting, hence we limited the modeling to an analytic phenomenological approximation (Section \ref{sec:SED}).
    Thus, we have no direct Lorentz factor estimates. 
    At very high redshift, $\Gamma\sim4-13$ have been measured or estimated up to now\footnote{More generally, bulk Lorentz factors range from 1 to 30 in the overall blazar population \citep[e.g.\ ][]{hovatta09}, but at $z>4$ a slightly narrower range has been observed up to now.}.
    Specifically, high-resolution radio observations allow for estimates of this parameter \citep[e,g,\ ][]{frey15,an20,spingola20,zhang20}, 
    while assumptions on its value are necessary to reliably model broad band SEDs and X-ray fluxes \citep[e.g.\ ][]{sbarrato12,sbarrato15,belladitta20, sbarrato22}. 
    The lowest $\Gamma$ values are generally obtained via radio observations, while the highest via broad-band modeling. 
    Therefore, we choose to calculate the densities for four relevant values: $\Gamma=4$, 8, 10, 13. 
    The core of the $\Gamma$ distribution lies between 8 and 13, and therefore we consider the related number densities in Figure \ref{Fig:Phi-zoom}, to compare the high-$z$ jetted AGN population with the comoving number density of massive quasars extracted from the most updated quasar luminosity function. 

    These results confirm that the fraction of jetted AGN significantly increases at very high redshift.
    We compare the derived jetted population in Table \ref{Table:Phi} with the comoving number density obtained from the luminosity function from \cite{shen20} with different bolometric luminosity cuts (Figure \ref{Fig:Phi-zoom}). 
    This comparison can be safely performed, because by selection the currently observed $z>4$ blazar population have rest frame optical-UV features completely consistent with non jetted quasars. 
    Their selection is performed by looking for clear big blue bump photometric features and subsequent spectroscopic confirmation, or by finding them directly in quasar catalogs. This currently ensures the selection of the most powerful FSRQs, with much redder synchrotron peaks than the average blazar population, and thus a very low contamination of the optical-UV continuum from the jet synchrotron emission \citep[][]{ghisellini09,ghisellini17,sbarrato13a,belladitta20}.
    Following \cite{volonteri11}, that suggested this approach for the first time, the main comparison is performed with the sources emitting $L_{\rm bol}\geq10^{47}$erg/s, that they assumed as corresponding to $>10^9M_\odot$ AGN accreting at the Eddington limit. 
    Depending on the Lorentz factor assumed, the comoving number density of massive jetted sources as traced by blazars can range from $\sim10\%$ of the corresponding value for the entire quasar population, up to its entirety.
   
    We consider $10^{47}$erg/s a solid limit to perform this comparison: (i) 75\% of sources in BLAZ4R have bolometric luminosities above that value, and (ii) if only the accretion disk emission is considered, sources with $L_{\rm bol}=10^{47}$erg/s are actually accreting below the Eddington limit. The bolometric luminosity in fact traces the overall AGN nuclear emission, including disk, Corona and IR torus. \cite{calderone13} derived that $L_{\rm bol}\simeq2\times2\cos\theta L_{\rm disk}$, and therefore the accretion emission is a factor 3-4 less than the bolometric one. This bolometric limit thus includes sources accreting at least at 25-30\% of the Eddington limit. 
    To explore more conservative definitions, we compare our results with $L_{\rm bol}>5\times10^{46}$erg/s as well. With this luminosity bin, the jetted fraction of the population ranges between $\sim5$ and 30\%.
    Nevertheless we stress that $5\times10^{46}$erg/s is below the median bolometric luminosity of our sample, and comparing BLAZ4R with the quasar luminosity function in this range translates in a very solid lower limit to the jetted fraction. 
    
    To summarise, in the first 1.5Gyr of the Universe, the most massive active SMBHs are hosted in significant fraction in jetted systems or, from a different point of view, their accretion and environment  likely triggers the launch of relativistic jets. 
    These results put significant and interesting constraints on the AGN formation and evolution theories, since they strongly affect the SMBH accretion processes, and mostly their evolution rates.

 \subsection{Central Engines}

    How such massive sources have evolved in less than 1.5~Gyr from the beginning of the Universe is still a widely open question.
    Not even accretion at the Eddington limit can justify their existence \citep[e.g.,][]{ghisellini13}. 
    A solid estimate of their accretion rate is thus necessary, in order to understand their accretion regime at the moment of the detection, and at most place a constraint on this evolutionary stage. 

\begin{figure}
   \centering
   \includegraphics[width=\hsize]{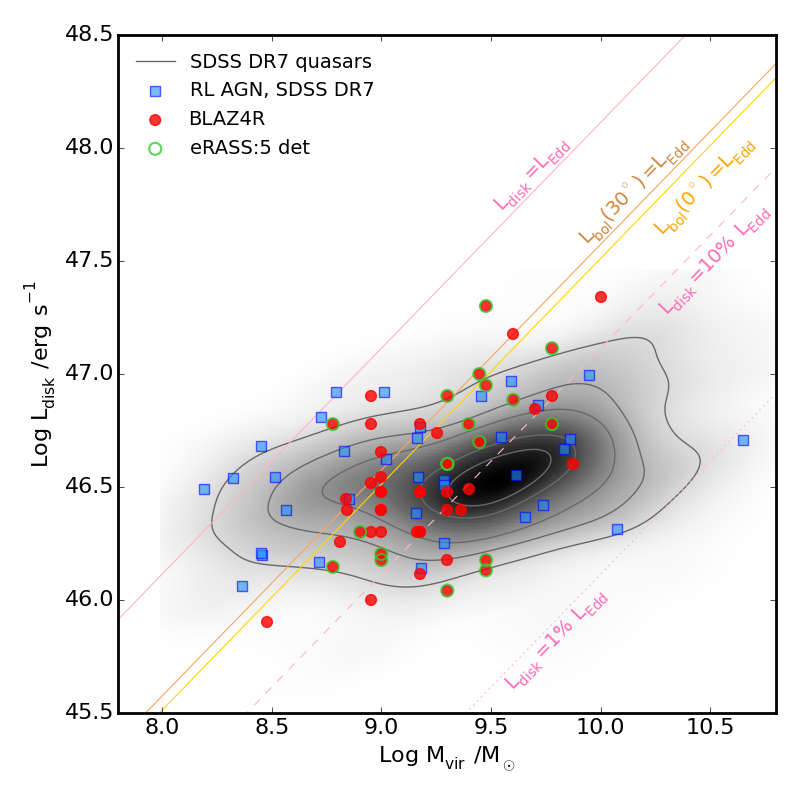}
      \caption{Disk luminosity as a function of black hole mass measures for our blazar sample 
               (filled red circles, circled in green if detected by eRASS:5), compared with $z>4$ SDSS DR7 quasars (grey contours) and $z>4$ SDSS DR7 radio-loud quasars (blue squares). 
               Notably, there is no significant difference in SMBH mass and accretion between confirmed blazars and the overall population traced by the SDSS survey. 
              }
         \label{Fig:mass-luminosity}
   \end{figure}

    Figure \ref{Fig:mass-luminosity} focuses on the accretion regime of $z>4$ AGN, by comparing the black hole mass and disk luminosity derived with the SED modeling of our sample with the overall $z>4$ SDSS quasar population, separated in radio-loud and radio-quiet/silent. 
    Particular attention must be given to the estimate of the disk luminosity values: data for the SDSS sources are directly derived from the 7$^{\rm th}$ SDSS Data Release quasar catalog \cite[SDSS DR7;][]{shen11}. 
    No direct estimate of the disk luminosity is given in that data release, but we derive its value from the bolometric luminosity, following 
    \citet{calderone13}.
    The bolometric luminosity is in fact calibrated by taking into account the overall nuclear emission  of an AGN, including disk, Coronal emission, and IR light from the obscuring torus. 
    Moreover, these components enter the $L_{\rm bol}$ definition with their intrinsic anisotropies. 
    Due to its geometry, the disk anisotropic emission can be described by an angular $2\cos\theta$ dependence. 
    Taking these effects into account, \citet[][]{calderone13} showed that $L_{\rm bol}\sim2\times2\cos\theta L_{\rm disk}$.
    We thus opt for deriving $L_{\rm disk}$ assuming an average viewing angle of $30^\circ$ for the overall SDSS quasar population, 
    and compare that value with the one derived by modeling the SEDs of our sources (see Section \ref{sec:SED}).

    Our blazar sample does not show significantly different nuclear features with respect to the overall $z>4$ population, 
    with a general fast accretion, but not preferentially super-critical. 
    In this regard, the effort put on deriving the accretion emission (bolometric vs.\ disk luminosity) is reflected in how the Eddington ratio is evaluated among the different samples we are considering. 
    For the sake of consistency with the literature, one should evaluate the Eddington ratio using the bolometric luminosity ($L_{\rm bol}/L_{\rm Edd}$). According to this definition, 13/51 of our sample would be super-Eddington, along with a fraction of the sources in the SDSS DR7.
    A more realistic definition of the Eddington ratio would rely on the radiative emission of the accretion process, i.e.\ the disk luminosity ($L_{\rm disk}/L_{\rm Edd}$). 
    According to the latter definition, Figure \ref{Fig:mass-luminosity} clearly shows that only one blazar would be classified as super-Eddington, along with three radio-loud sources and a low-significant tail of the SDSS-DR7 quasars ($2.4\%$). 
    Nonetheless, most sources in our sample appear to accrete at very high Eddington ratios, most of them with $L_{\rm disk}>10\%L_{\rm Edd}$. 

    It is worth noticing that such a fast accretion is however not enough to justify the measured black hole masses at $z>4$. 
    Not even an Eddington-limited accretion is sufficient: many works have shown that under the reasonable assumptions of a $\sim100M_\odot$ black hole seed at $z\sim20$ and a continuous accretion at the Eddington rate, a $\sim10^9M_\odot$ active black hole can form by $z\sim4$ only in the case of a strictly non-spinning black hole, and the amount of time is barely sufficient even in this optimistic case \citep[see e.g.\ ][]{ghisellini13,banados15,sbarrato21}. The sources observed at $z\sim5-7$ cannot instead be justified. 
    A more reasonable assumption of a maximally spinning black hole, often invoked in the presence of a relativistic jet, would require at least 3.1~Gyr to form a $10^9M_\odot$ black hole. 
    The observed accretion rates and related Eddington ratios are even slower than the simplistic Eddington-limited assumption: other paths must be explored. 
    Super-Eddington accretion would clearly be the cleaner solution, and lately new classes of sources that could be associated with super-critical evolution are being explored \citep[Little Red Dots, see e.g.\ ][]{matthee24}.
    However, the large space density of sub-Eddington, yet massive jetted AGN implies that the super-Eddington episodes must be fast, and able to form a fully grown accreting SMBH, consistent with what is observed at lower $z$ \citep{alzati25}.

%

\section{Conclusions}

In the last 25 years, our knowledge of high-$z$ AGN has enormously increased. 
Since the early 2000s, we understood that they strongly challenge the current black hole formation and early evolution models, reaching too high masses in very short timescales ($>10^9M_\odot$ in less than 2 billion, or even 1 billion years). 
Their common overobscuration, and the quenching of extended radio emission makes jetted sources difficult to detect at $z>4$, and therefore blazars have been used in the last 10 years as effective tracers of the overall jetted population.
This has lead to the conclusion that jetted AGN are in fact over-represented among active massive SMBHs, challenging the black hole formation paradigm at first, and then suggesting a role for relativistic jets in the fast assembly of the earliest AGN.

In this work, we focused on the X-ray emission of $z>4$ blazars and blazar candidates, along with their multiwavelength features and radio counterparts, in order to collect all the known aligned jetted sources in this redshift bin and study them collectively.
We collected all 53 known blazars from the literature, along with 11 reliable blazar candidates, not confirmed by X-ray features, then cross-correlated them with the most recent data from eROSITA, i.e.\ the 4 complete scans and the portion of the sky covered by the beginning of their 5th (eRASS:5). 
17 of the originally known blazars are detected by eRASS:5, and so is one of the candidate. We classified the latter as a blazar thanks to broad-band SED modeling, with the crucial help of its bright, hard X-ray spectrum. 
The final sample of 54 $z>4$ blazars defines the starting point of  BLAZ4R, a living catalog of all the known $z>4$, complete with SED characterization, estimates of black hole masses and accretion luminosity, radio features where available.

To understand the limitations of eRASS:5, we derived upper limits for the undetected sources, and obtained detailed X-ray spectra of the 8 brightest blazars detected. 
4 sources had more than 80 counts across the 5 scans, and this allowed us to check for temporal variability, that unfortunately resulted not being significant for any of them. 
We cross-matched  BLAZ4R and the candidates with many public radio catalog, finding a variety of radio spectral features across the extended sample (44 flat, 12 peaked, 2 convex and 4 flat+steep radio spectra), and for some of the sources hints of variability. 
We modeled the  BLAZ4R broad-band SEDs with a phenomenological model \citep{ghisellini17}, obtaining ranges of possible SED profiles, all characterized by typical blazar features, and more importantly the extreme Compton Dominance already observed for $z>4$ blazars. 
This profile confirms the sample to be composed by the brightest end of possible jet powers distribution in the general family of blazars.

 BLAZ4R is the most complete description of known $z>4$ to date, and we plan to keep it updated with new discoveries and classifications. 
This will allow us to keep up to date and refine the comoving number densities of  massive jetted AGN at $z>4$, in order to compare them with the space densities of the overall quasar population.
Currently, the jetted fraction appears to be significantly larger than the average $\sim10\%$ observed at lower redshift, up to the entirety of the quasar population if particularly powerful jets are assumed. This suggests a relevant role for relativistic jets in the evolution of the most massive SMBHs in the first 2 billion years of cosmic history. 
Finally, we focused on the nuclear features of our sample with what derived in the literature for the general $z>4$ quasar population (both radio-loud and quiet), finding them to be in line with the overall behavior. This somehow enhances the issue of their fast early formation: no significantly super-Eddington source is found, while their masses are on the extreme end of the distribution.

Our work ultimately highlights the significance of relativistic jets in the early formation and evolution of the earliest, most massive supermassive black holes. 
We will keep on monitoring the population, in search of signatures of fast accretion, simultaneous to the jet activity, or even right before their launch. 
We cannot exclude that the jet over-abundance among the most massive sources might be a signature of past super-critical activity, that somehow would trigger the launch of powerful jets.

\begin{acknowledgements}
    S.H. is partly supported by the German Science Foundation (DFG grant numbers WI 1860/14-1 and 434448349). The National Radio Astronomy Observatory is a facility of the U.S. National Science Foundation operated under cooperative agreement by Associated Universities, Inc.
    This paper includes archived data obtained through the CSIRO ASKAP Science Data Archive, CASDA (http://data.csiro.au). This scientific work uses data obtained from Inyarrimanha Ilgari Bundara / the Murchison Radio-astronomy Observatory. We acknowledge the Wajarri Yamaji People as the Traditional Owners and native title holders of the Observatory site. CSIRO’s ASKAP radio telescope is part of the Australia Telescope National Facility (https://ror.org/05qajvd42). Operation of ASKAP is funded by the Australian Government with support from the National Collaborative Research Infrastructure Strategy. ASKAP uses the resources of the Pawsey Supercomputing Research Centre. Establishment of ASKAP, Inyarrimanha Ilgari Bundara, the CSIRO Murchison Radio-astronomy Observatory and the Pawsey Supercomputing Research Centre are initiatives of the Australian Government, with support from the Government of Western Australia and the Science and Industry Endowment Fund.
    The National Radio Astronomy Observatory is a facility of the U.S. National Science Foundation operated under cooperative agreement by Associated Universities, Inc.
    Part of this work is based on archival data, software or online services provided by the Space Science Data Center - ASI.
\end{acknowledgements}

\bibliographystyle{aa} 
\bibliography{BIB} 

\appendix

\section{SED modeling results}

\begin{figure*}
   \centering
   \includegraphics[width=0.4\hsize]{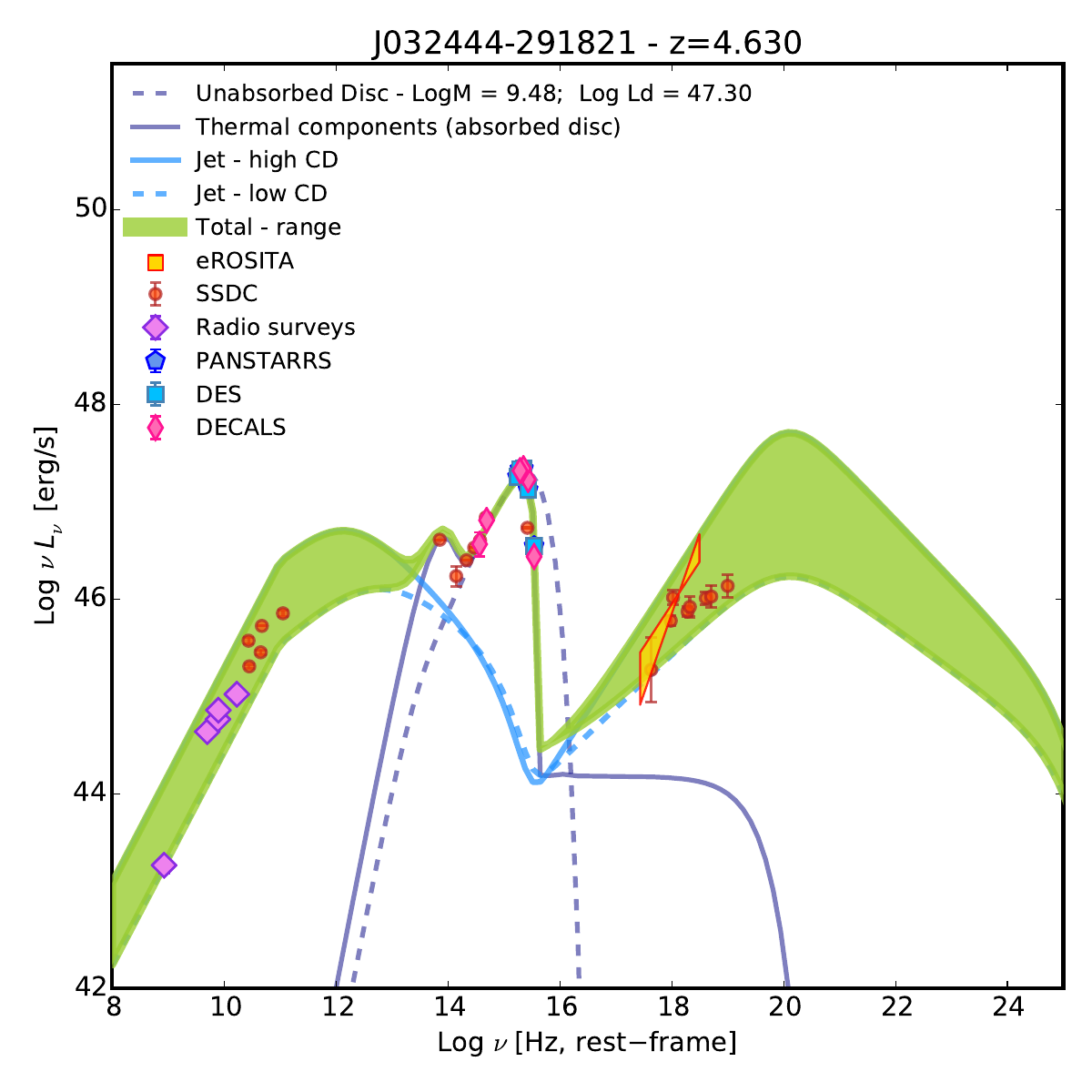}
   \includegraphics[width=0.4\hsize]{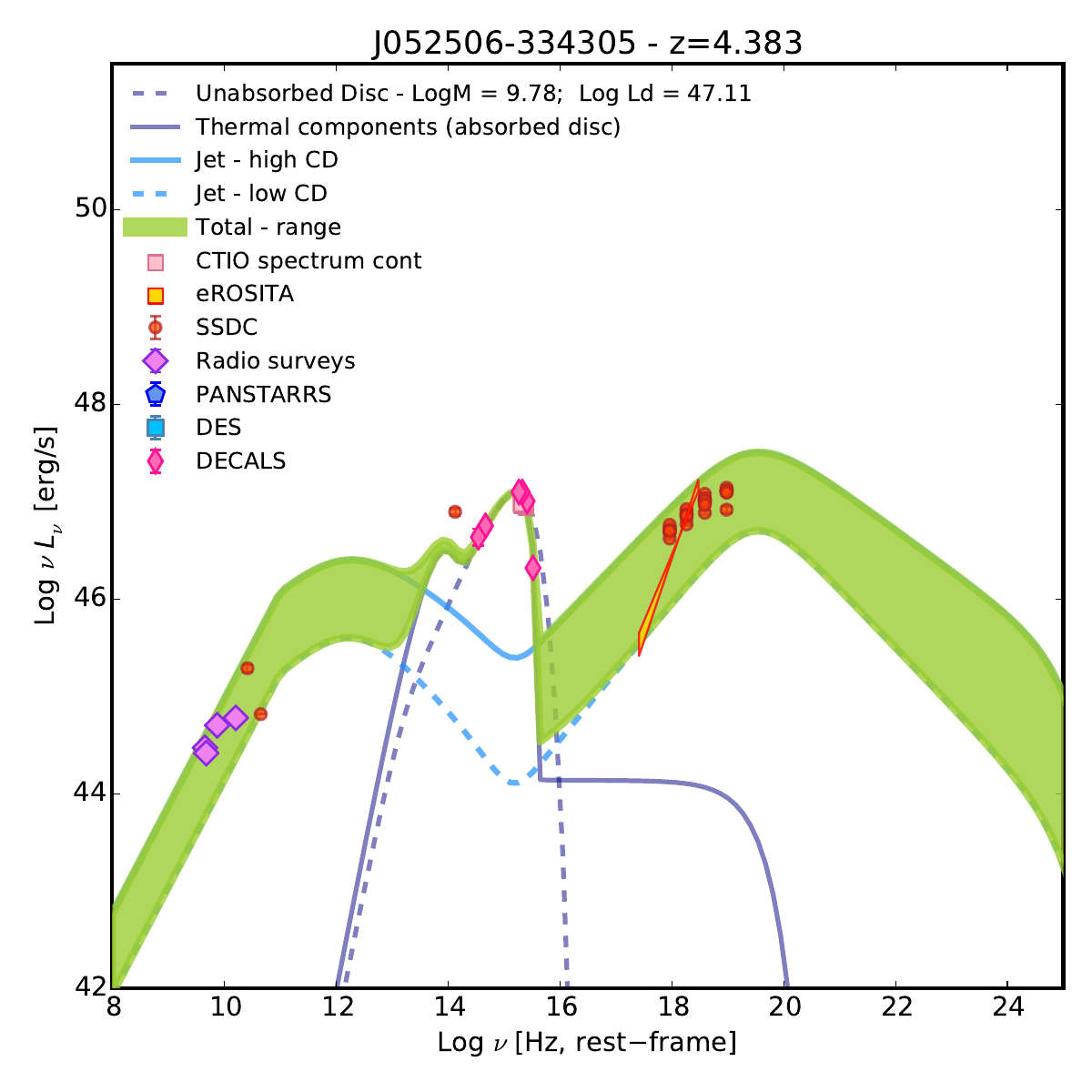}
   \includegraphics[width=0.4\hsize]{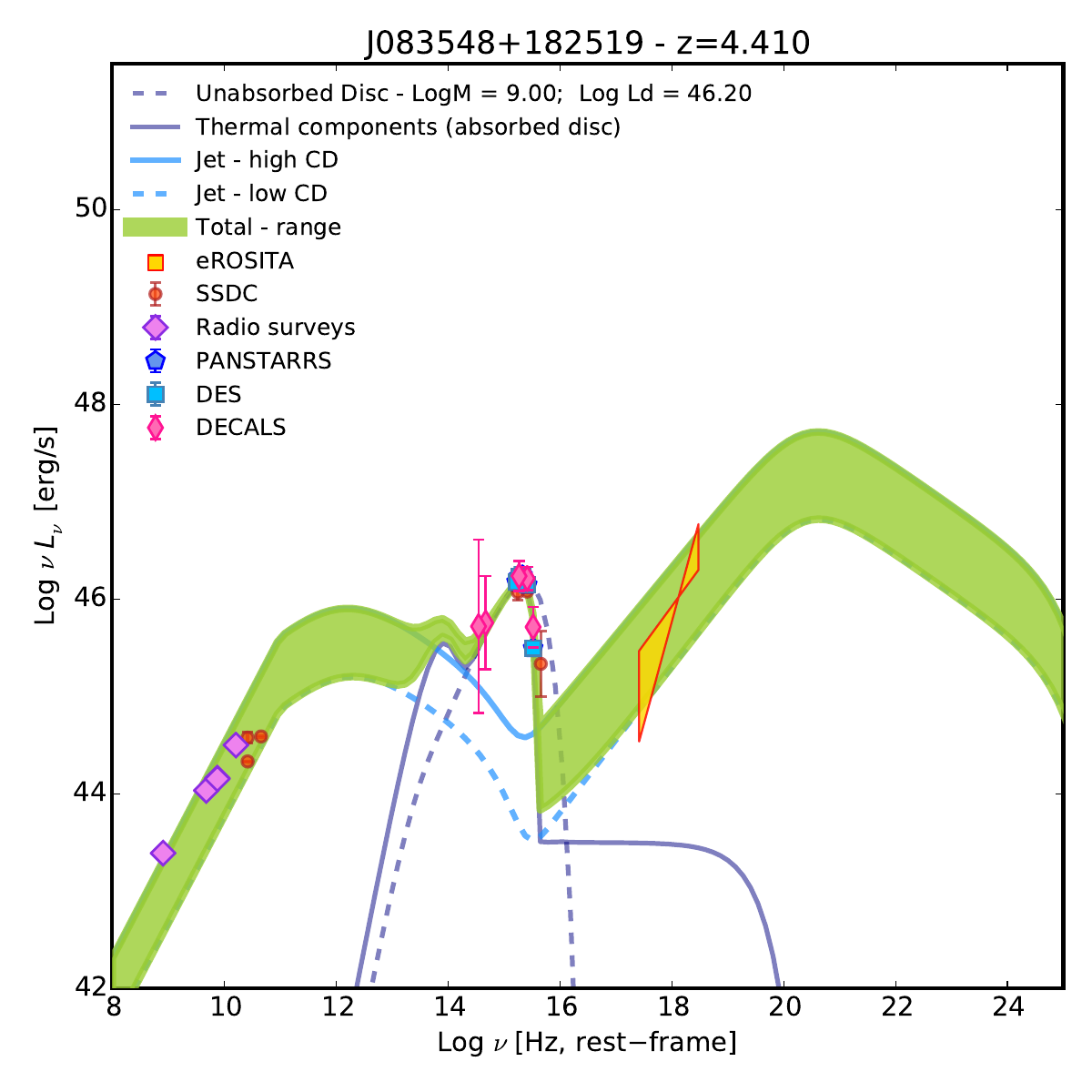}
   \includegraphics[width=0.4\hsize]{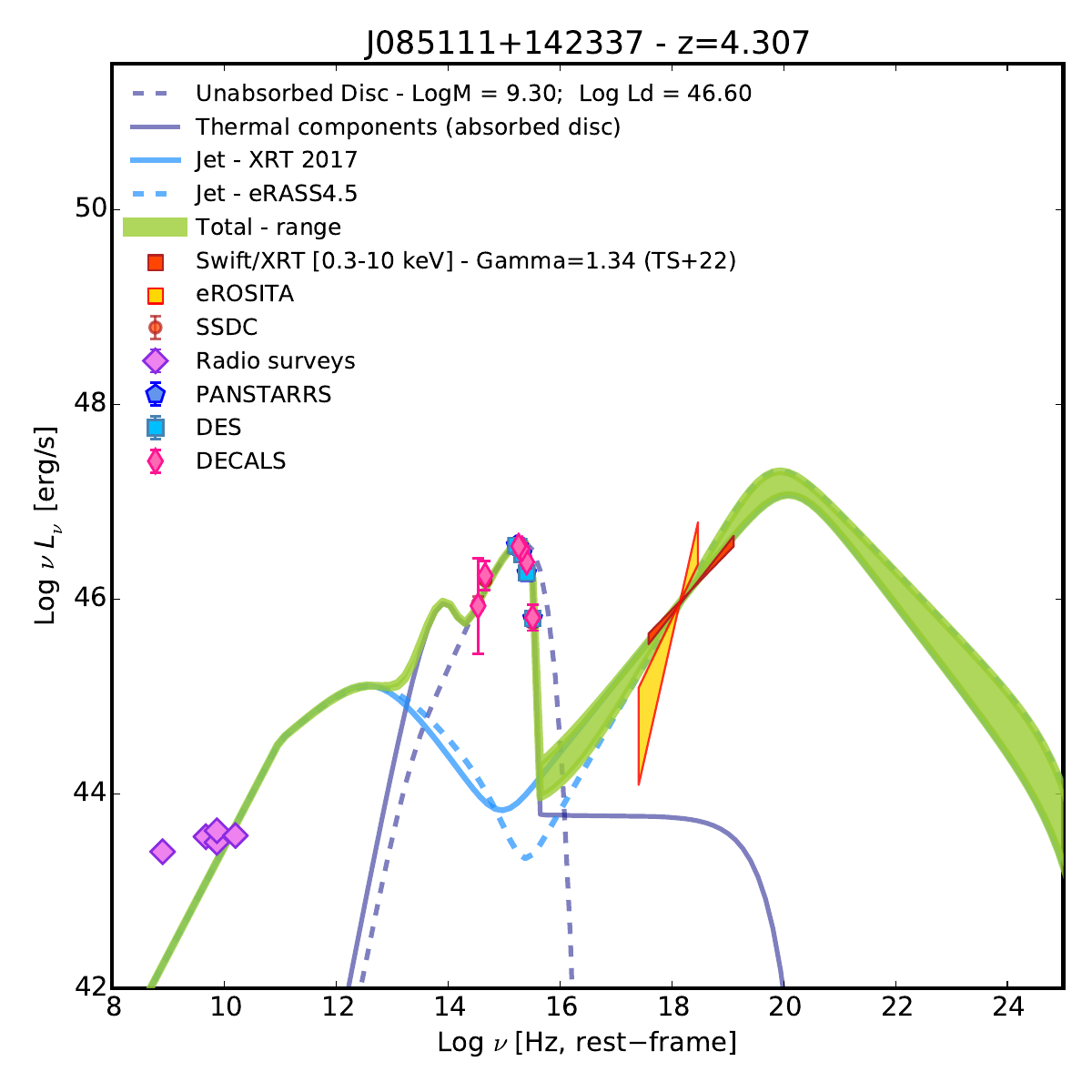}
      \caption{SED modeling of the brightest sources in eRASS:5.
                All panels include data collected from public catalogs (SSDC, PANSTARRS, DES, DECALS, as labelled), radio surveys explored in the original BLAZ4R paper, possible X-ray data from the literature and eROSITA data. 
                Blue solid lines show the absorbed thermal components (accretion disk, torus and X-ray Corona), blue dashed lines the unabsorbed disk, solid and dashed light blue lines are the two limiting jet SED solutions (lower and higher respectively). The green shaded regions highlight the range of possible total SEDs that would reliably describe the sources.
              }
         \label{Fig:SED1}
   \end{figure*}

\begin{figure*}
   \centering
   \includegraphics[width=0.4\hsize]{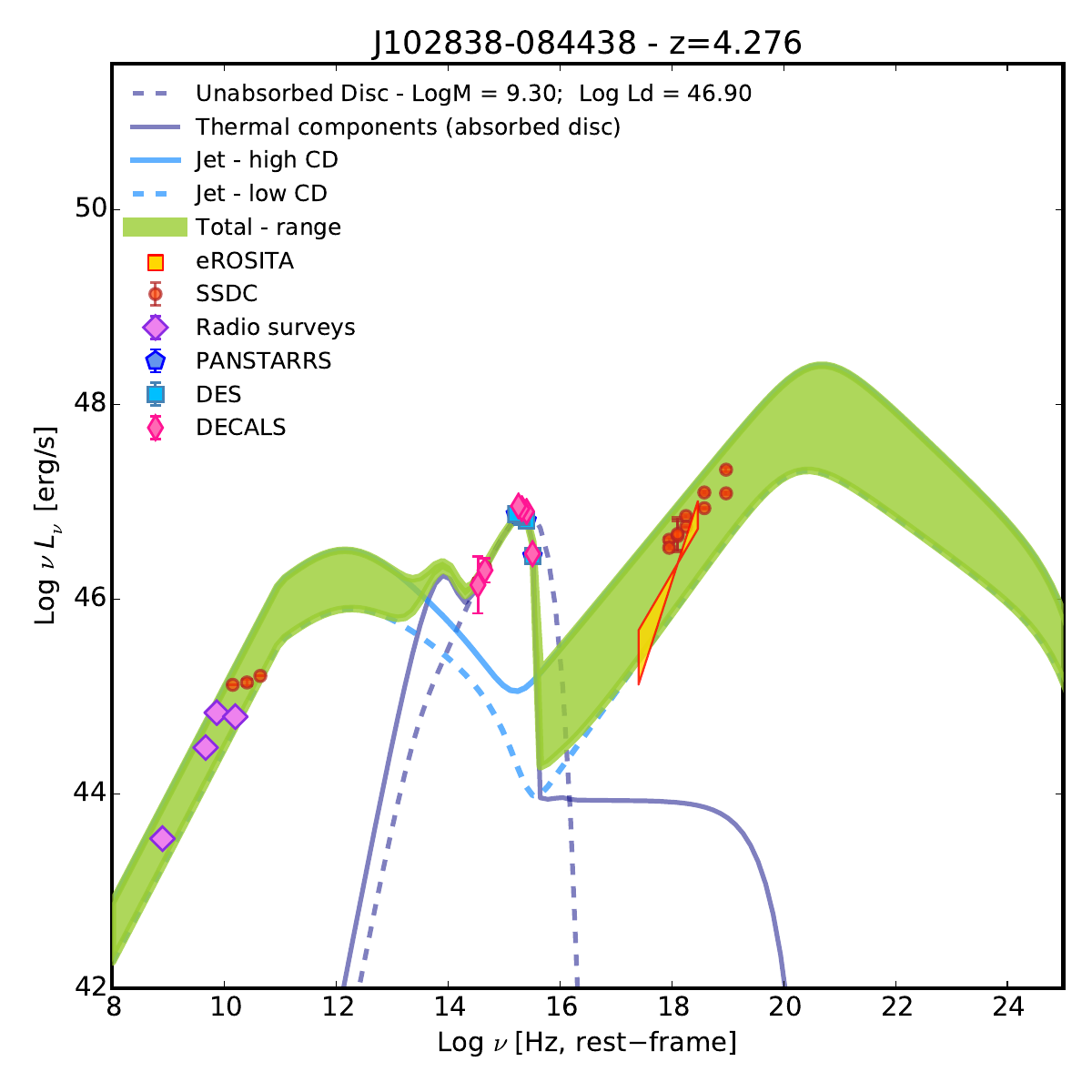}
   \includegraphics[width=0.4\hsize]{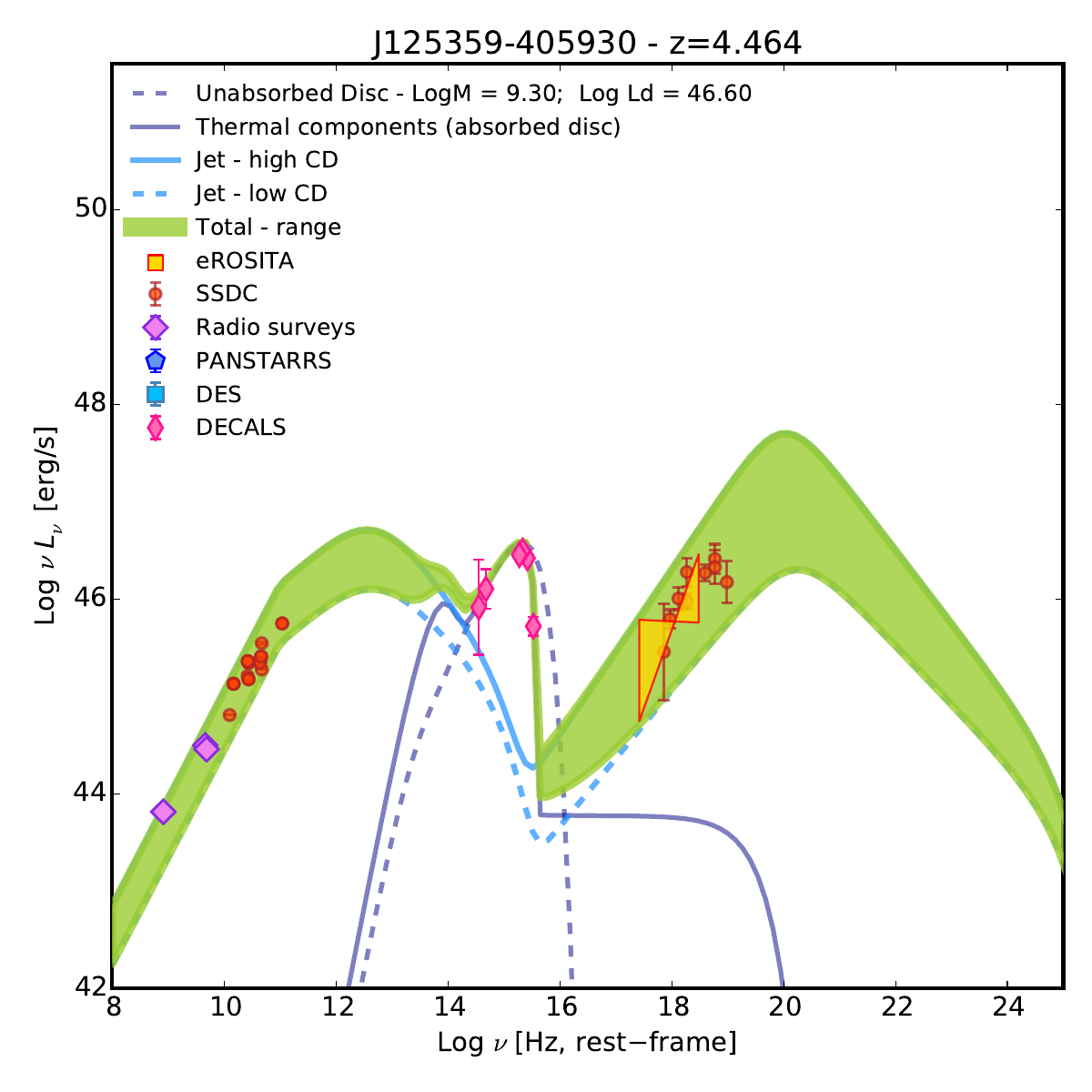}
   \includegraphics[width=0.4\hsize]{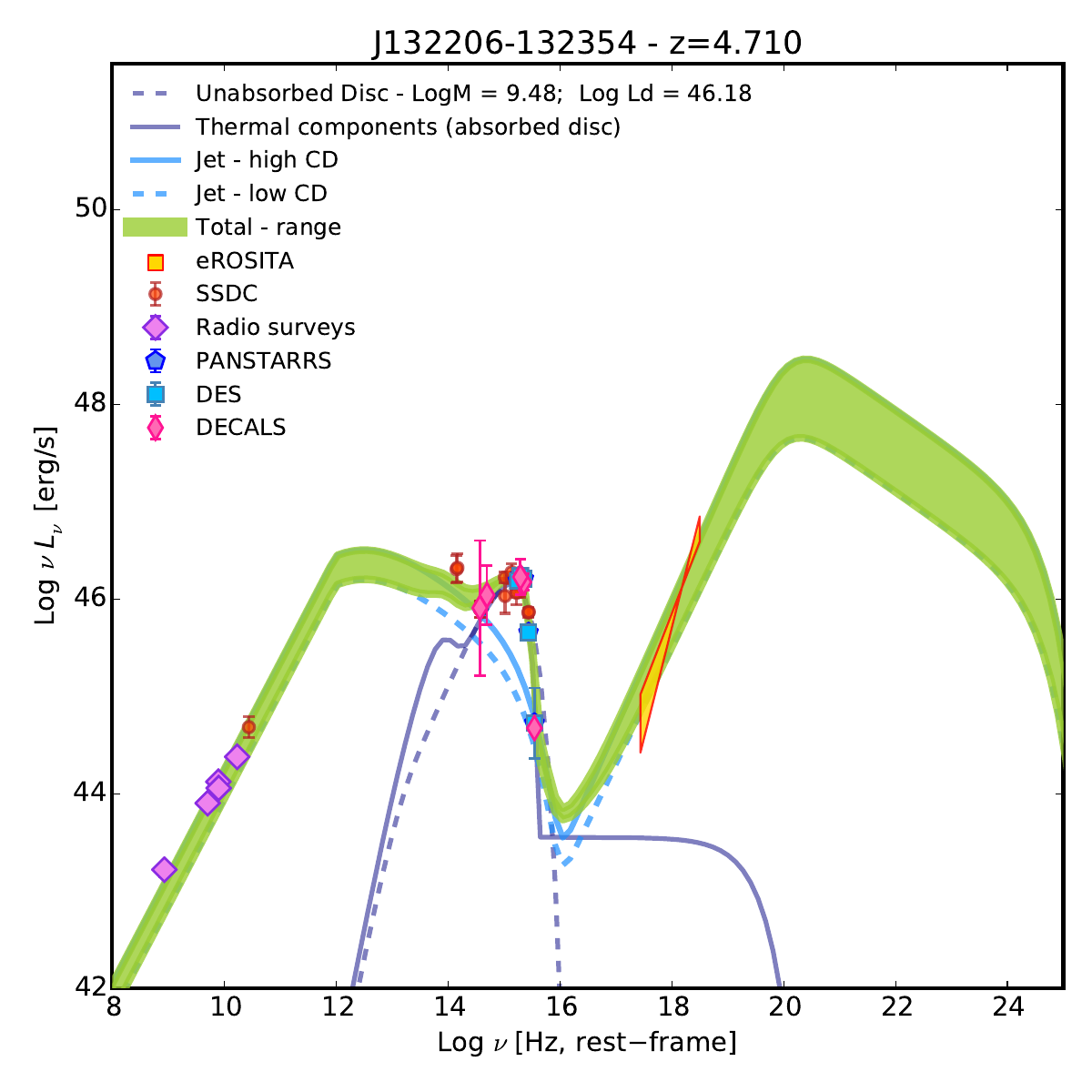}
   \includegraphics[width=0.4\hsize]{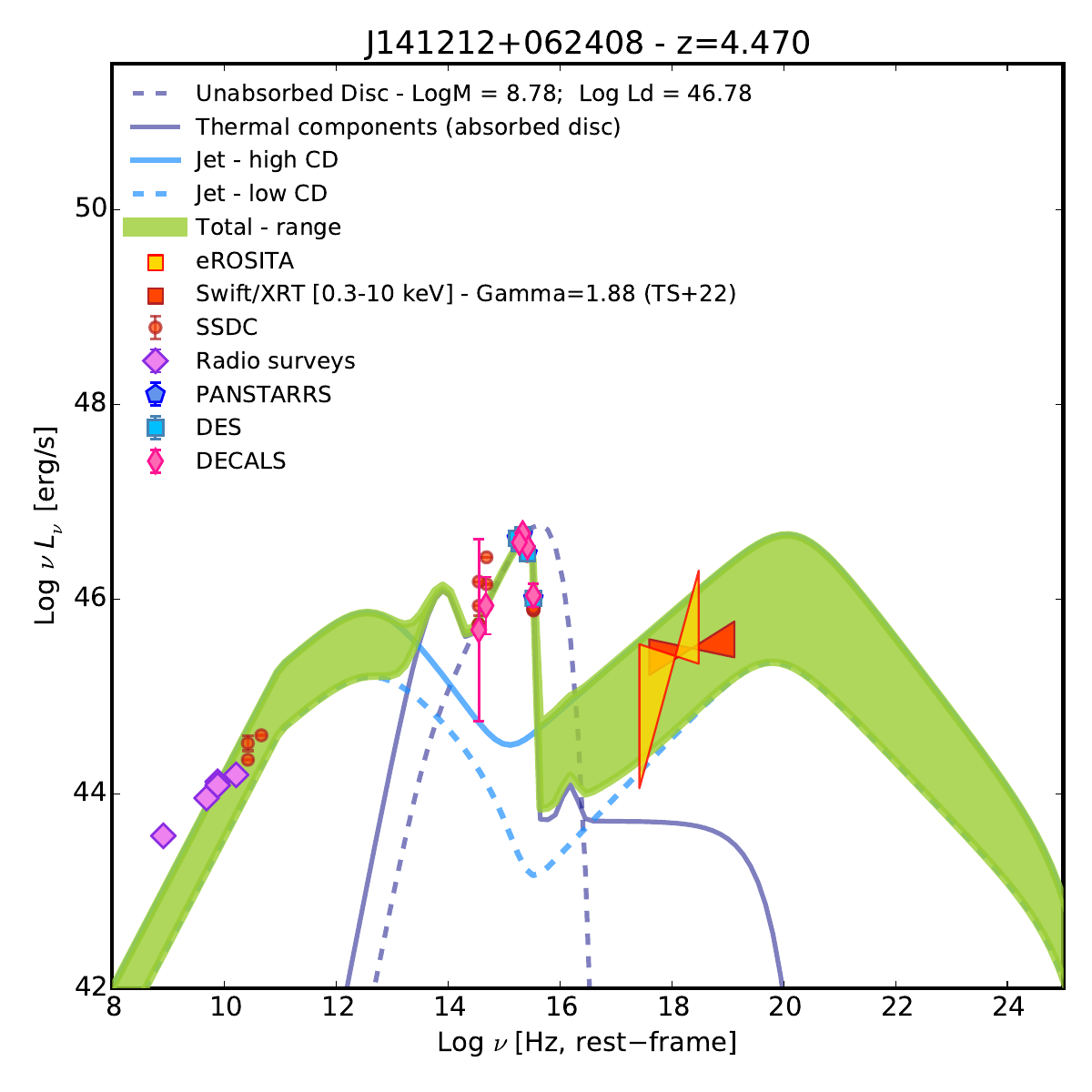}
      \caption{SED modeling of the brightest sources in eRASS:5. See caption of Figure \ref{Fig:SED1}
              }
         \label{Fig:SED2}
   \end{figure*}

We applied the phenomenological model by \citep{ghisellini17} to describe the broad-band emission of the sources included in BLAZ4R. 
An $\alpha$-disk model is also considered to take into account the thermal emission from the accretion disk, and derive mass and accretion luminosity of the central SMBH. 
All the details of the modeling are reported in Section \ref{sec:SED}. 
Since no high-energy data are available for the sample, the blazar jet SEDs could not significantly constrained: for this reason, we derive a range of validity for the parameters that define the model. 
The parameters related to the accreting system (i.e.\ mass and disk luminosity) are better constrained, but we chose not to derive any uncertainty on their values, since their big blue bumps are significantly affected by intervening Ly$\alpha$ cloud absorption and the needed correction add a further level of uncertainty. As thoroughly discussed in \citet{sbarrato13a} and \citet{calderone13}, a reasonable level of uncertainty for this approach is of a factor of 2 over the black hole mass, i.e.\ comparable or even smaller than the typical precision of the commonly used virial model \citep[at least a factor 3-4, see][]{vestergaard06}.
All results are reported in Tables \ref{Table:SED-results1} and \ref{Table:SED-results2}.
All SEDs are available in the dedicated BLAZ4R website\footnote{\url{https://blaz4r.brera.inaf.it/}}, while Figures \ref{Fig:SED1} and \ref{Fig:SED2} show those of the 8 brightest eROSITA sources of BLAZ4R.
For these blazars we could obtain eROSITA X-ray spectra, and the IC component is thus better constrained. 

%
%

%
\begin{sidewaystable*}
\caption{\tiny Results of the SED modeling, performed by following \cite{ghisellini17}.         
         }
\label{Table:SED-results1}      
\centering  
\tiny
\begin{tabular}{ccccccccccccccc}
\hline \hline
  Name & $z$ & Log$M/M_\odot$  & Log$L_{\rm d}$  & $\alpha_1$ & $\alpha_2$ & $\alpha_3$ & Log$\nu_{\rm t}$ &  Log$\nu_{\rm S}$ &  Log$\nu_{\rm C}$ &  Log$\nu_{\rm cut,S}$ &   Log$\nu_{\rm cut,C}$ &   Log$\nu_{\rm S}L(\nu_{\rm S})$ & Log(CD) & model \\
   & & & [erg/cm$^2$/s]  &  &  &  & [Hz] &  [Hz] &  [Hz] &  [Hz] &   [Hz] & [erg/cm$^2$/s] & & \\
\hline
  J001115+144601 & 4.96 & 9.60 & 47.18 & 0.4 & 1.55 & 0.6 & 11.0 & 13.5 & 21.5 & 15.0 & 25.0 & 46.4 & 1.15 & high \\
                 &      &      &       & 0.4 & 1.42 & 0.65 & 11.0 & 13.5 & 21.2 & 15.0 & 25.0 & 45.8 & 1.1 & low\\
  J012202+030951 & 4.0 & 9.30 & 46.18 & 0.5 & 1.55 & 0.08 & 11.0 & 12.4 & 21.4 & 16.3 & 26.48 & 46.2 & 3.5 & high \\
                 &      &      &       & 0.5 & 1.42 & 0.24 & 11.0 & 12.4 & 21.4 & 16.0 & 27.48 & 45.6 & 2.5 & low\\
  J012714-445445 & 4.96 & 8.78 & 46.15 & 0.4 & 1.75 & 0.4 & 11.5 & 12.0 & 20.5 & 15.0 & 25.0 & 45.9 & 1.5 & high \\
                 &      &      &       & 0.4 & 1.8 & 0.5 & 11.5 & 12.0 & 20.0 & 15.0 & 24.0 & 45.3 & 1.0 & low \\
  J013127-032100 & 5.18 & 10.0 & 47.34 & 0.5 & 1.5 & 0.45 & 10.8 & 11.6 & 19.3 & 15.0 & 22.0 & 45.6 & 0.9 & high \\
                 &      &      &       & 0.5 & 1.45 & 0.45 & 11.0 & 11.5 & 19.0 & 15.0 & 22.0 & 45.0 & 1.0 & low\\
  J014132-542749 & 5.0 & 9.00 & 46.18 & 0.5 & 1.7 & 0.65 & 11.5 & 12.4 & 20.5 & 15.0 & 25.0 & 46.4 & -0.3 & high \\
                 &      &      &       & 0.5 & 1.8 & 0.7 & 11.5 & 12.4 & 20.0 & 15.0 & 24.0 & 45.8 & -2.0 & low\\
  J020228-170827 & 5.57 & 9.45 & 47.0 & 0.5 & 1.75 & 0.3 & 11.0 & 12.0 & 20.8 & 15.0 & 25.0 & 46.0 & 1.6 & high \\
                 &      &      &       & 0.5 & 1.9 & 0.5 & 11.0 & 12.5 & 20.8 & 15.0 & 25.0 & 45.0 & 1.0 & low\\
  J020916-562650 & 5.606 & 8.90 & 46.30 & 0.5 & 1.6 & 0.4 & 11.5 & 12.0 & 21.0 & 15.0 & 25.0 & 45.9 & 1.5 & high \\
                 &      &      &       & 0.5 & 1.45 & 0.8 & 11.5 & 12.0 & 20.8 & 15.0 & 24.0 & 45.3 & 0.1 & low\\
  J021043-001818 & 4.77 & 9.30 & 46.60 & 0.5 & 1.6 & 0.4 & 11.5 & 12.5 & 20.5 & 15.0 & 25.0 & 45.9 & 1.1 & high \\
                 &      &      &       & 0.5 & 1.45 & 0.8 & 11.5 & 12.5 & 20.0 & 15.0 & 24.0 & 45.0 & -0.1 & low\\
  J025758+433837 & 4.07 & 9.00 & 46.40 & 0.5 & 1.8 & 0.2 & 11.5 & 12.3 & 20.5 & 15.0 & 25.0 & 47.0 & 1.5 & high \\
                 &      &      &       & 0.5 & 1.45 & 0.5 & 11.5 & 12.5 & 20.0 & 15.0 & 24.0 & 46.1 & 0.01 & low\\
  J030437+004653 & 4.305 & 9.00 & 46.40 & 0.5 & 1.7 & 0.6 & 11.5 & 12.0 & 20.5 & 15.0 & 25.0 & 45.9 & 0.4 & high \\
                 &      &      &        & 0.4 & 1.7 & 0.9 & 11.5 & 12.3 & 19.0 & 15.0 & 24.0 & 45.5 & -0.73 & low\\  
  J030947+271757 & 6.1 & 9.16 & 46.30 & 0.5 & 1.8 & 0.55 & 11.5 & 12.5 & 22.0 & 15.0 & 25.0 & 46.3 & 1.5 & high \\
                 &      &      &       & 0.5 & 1.5 & 0.7 & 11.5 & 12.5 & 22.0 & 15.0 & 24.0 & 45.2 & 1.1 & low\\
  J032214-184117 & 6.09 & 9.18 & 46.30 & 0.5 & 1.5 & 0.7 & 11.5 & 12.0 & 19.5 & 15.0 & 21.0 & 45.0 & 0.9 & high \\
                 &      &      &       & 0.5 & 1.5 & 0.8 & 11.5 & 12.0 & 18.5 & 15.0 & 21.0 & 44.5 & 0.4 & low \\   
  J032444-291821 & 4.63 & 9.477 & 47.30 & 0.5 & 1.6 & 0.1 & 11.0 & 12.2 & 20.0 & 15.0 & 25.0 & 46.7 & 1.0 & high \\
                 &      &      &       & 0.5 & 1.42 & 0.45 & 11.0 & 12.7 & 20.0 & 15.0 & 25.0 & 46.1 & 0.13 & low\\
  J041009-013919 & 6.995 & 8.84 & 47.00 & 0.5 & 1.8 & 0.3 & 11.5 & 12.5 & 22.0 & 15.0 & 25.0 & 46.0 & 1.8 & high \\
                 &      &      &       & 0.5 & 1.5 & 0.5 & 11.5 & 12.5 & 21.5 & 15.0 & 24.0 & 45.2 & 1.3 & low\\
  J052506-334305 & 4.383 & 9.78 & 47.11 & 0.5 & 1.42 & 0.4 & 11.0 & 12.2 & 19.4 & 15.0 & 25.0 & 46.4 & 1.1 & high \\
                 &      &      &       & 0.5 & 1.6 & 0.3 & 11.0 & 12.3 & 19.5 & 15.0 & 25.0 & 45.6 & 1.1 & low\\
  J081333+350810 & 4.922 & 9.48 & 46.95 & 0.5 & 1.7 & 0.65 & 11.5 & 12.0 & 20.5 & 15.0 & 25.0 & 46.2 & 0.05 & high \\
                 &      &      &       & 0.5 & 1.6 & 0.9 & 11.5 & 12.0 & 20.0 & 15.0 & 24.0 & 45.2 & -1.5 & low\\
  J083548+182519 & 4.41 & 9.00 & 46.20 & 0.5 & 1.4 & 0.3 & 11.0 & 12.1 & 20.4 & 15.0 & 25.0 & 45.9 & 1.8 & high \\
                 &      &      &       & 0.5 & 1.4 & 0.25 & 11.0 & 12.2 & 20.4 & 15.0 & 25.0 & 45.2 & 1.6 & low\\
  J083946+511202 & 4.4 & 9.78 & 46.90 & 0.5 & 1.7 & 0.5 & 11.5 & 11.8 & 20.5 & 14.0 & 25.0 & 46.0 & 1.0 & high \\
                 &      &      &       & 0.5 & 1.7 & 0.6 & 11.5 & 12.0 & 20.5 & 14.0 & 24.0 & 45.2 & 1.0 & low\\
  J085111+142337 & 4.307 & 9.30 & 46.60 & 0.5 & 1.75 & 0.28 & 11.0 & 12.7 & 20.1 & 15.0 & 25.0 & 45.1 & 1.97 & high \\
                 &      &      &       & 0.5 & 1.6 & 0.005 & 11.0 & 12.7 & 19.8 & 15.0 & 25.0 & 45.1 & 2.2 & low\\
  J090132+161506 & 5.63 & 9.30 & 46.48 & 0.5 & 1.6 & 0.4 & 11.5 & 12.0 & 21.0 & 15.0 & 25.0 & 45.5 & 1.5 & high \\
                 &      &      &       & 0.5 & 1.45 & 0.9 & 11.5 & 12.5 & 20.8 & 15.0 & 24.0 & 44.6 & -0.3 & low\\
  J090630+693030 & 5.47 & 8.95 & 46.90 & 0.5 & 1.6 & 0.35 & 11.5 & 12.3 & 23.3 & 14.0 & 25.0 & 46.9 & 2.5 & high \\
                 &      &      &       & 0.5 & 1.7 & 0.5 & 11.5 & 12.2 & 22.0 & 14.0 & 24.5 & 46.1 & 1.0 & low\\
  J091824+063653 & 4.22 & 9.78 & 46.78 & 0.5 & 1.6 & 0.55 & 11.0 & 12.7 & 20.4 & 15.0 & 25.0 & 46.0 & 0.8 & high \\
                 &      &      &       & 0.5 & 1.42 & 0.65 & 11.0 & 12.7 & 20.0 & 15.0 & 25.0 & 45.35 & 0.1 & low\\
  J094004+052630 & 4.5 & 9.48 & 46.13 & 0.5 & 1.4 & 0.3 & 11.0 & 12.2 & 20.6 & 15.0 & 25.0 & 45.6 & 1.7 & high \\
                 &      &      &       & 0.5 & 1.4 & 0.4 & 11.0 & 12.2 & 20.6 & 15.0 & 25.0 & 45.2 & 0.9 & low\\
  J101335+281119 & 4.75 & 8.95 & 46.00 & 0.5 & 1.6 & 0.35 & 11.5 & 12.3 & 22.3 & 14.0 & 24.5 & 45.8 & 2.0 & high \\
                 &      &      &       & 0.5 & 1.7 & 0.5 & 11.5 & 12.2 & 22.0 & 14.0 & 24.5 & 45.1 & 1.0 & low\\
  J102107+220904 & 4.26 & 9.30 & 46.04 & 0.5 & 1.5 & 0.3 & 11.0 & 12.0 & 20.6 & 15.0 & 25.0 & 46.0 & 1.3 & high \\
                 &      &      &       & 0.5 & 1.5 & 0.65 & 11.0 & 12.2 & 20.6 & 15.0 & 25.0 & 45.4 & -0.2 & low\\
  J102623+254259 & 5.3 & 9.45 & 46.70 & 0.5 & 1.75 & 0.11 & 11.0 & 12.5 & 20.2 & 15.0 & 25.0 & 46.4 & 1.4 & high \\
                 &      &      &       & 0.5 & 1.6 & 0.25 & 11.0 & 12.7 & 20.2 & 15.0 & 25.0 & 46.0 & 0.5 & low\\
  J102838-084438 & 4.276 & 9.30 & 46.90 & 0.5 & 1.55 & 0.3 & 11.0 & 12.2 & 20.6 & 15.0 & 25.0 & 46.5 & 1.9 & high \\
                 &      &      &       & 0.5 & 1.42 & 0.2 & 11.0 & 12.2 & 20.2 & 15.0 & 25.0 & 45.9 & 1.4 & low\\
\hline                                   
\end{tabular}
\end{sidewaystable*}
\begin{sidewaystable*}
\caption{\tiny Following from \ref{Table:SED-results1}         
         }
\label{Table:SED-results2}      
\centering  
\tiny
\begin{tabular}{ccccccccccccccc}
\hline \hline
  Name & $z$ & Log$M/M_\odot$  & Log$L_{\rm d}$  & $\alpha_1$ & $\alpha_2$ & $\alpha_3$ & Log$\nu_{\rm t}$ &  Log$\nu_{\rm S}$ &  Log$\nu_{\rm C}$ &  Log$\nu_{\rm cut,S}$ &   Log$\nu_{\rm cut,C}$ &   Log$\nu_{\rm S}L(\nu_{\rm S})$ & Log(CD) & model \\
   & & & [erg/cm$^2$/s]  &  &  &  & [Hz] &  [Hz] &  [Hz] &  [Hz] &   [Hz] & [erg/cm$^2$/s] & & \\
\hline
  J114657+403708 & 5.005 & 9.70 & 46.85 & 0.5 & 1.6 & 0.4 & 11.5 & 12.3 & 22.3 & 14.0 & 24.5 & 45.8 & 2.3 & high \\
                 &      &      &       & 0.5 & 1.7 & 0.5 & 11.5 & 12.2 & 22.0 & 14.0 & 24.5 & 45.1 & 2.0 & low\\
  J115503-310758 & 4.3 & 9.40 & 46.78 & 0.5 & 1.75 & 0.55 & 11.0 & 12.5 & 20.2 & 15.0 & 25.0 & 46.5 & 0.3 & high \\
                 &      &      &       & 0.5 & 1.6 & 0.55 & 11.0 & 12.7 & 20.2 & 15.0 & 25.0 & 45.7 & 0.0 & low\\
  J123142+381658 & 4.137 & 8.85 & 46.40 & 0.5 & 1.6 & 0.5 & 11.5 & 12.3 & 22.3 & 14.0 & 24.5 & 45.4 & 2.2 & high \\
                 &      &      &       & 0.5 & 1.7 & 0.5 & 11.5 & 12.2 & 22.0 & 14.0 & 24.5 & 45.0 & 1.5 & low\\
  J123503-000331 & 4.723 & 8.95 & 46.52 & 0.5 & 1.6 & 0.5 & 11.5 & 11.6 & 22.3 & 14.0 & 24.5 & 45.3 & 2.2 & high \\
                 &      &      &       & 0.5 & 1.7 & 0.5 & 11.5 & 11.8 & 22.0 & 14.0 & 24.5 & 44.7 & 1.5 & low\\
  J125359-405930 & 4.464 & 9.30 & 46.60 & 0.5 & 1.75 & 0.15 & 11.0 & 12.7 & 20.0 & 15.0 & 25.0 & 46.7 & 1.0 & high \\
                 &      &      &       & 0.5 & 1.6 & 0.3 & 11.0 & 12.7 & 20.2 & 15.0 & 25.0 & 46.1 & 0.2 & low\\
  J130940+573309 & 4.28 & 8.95 & 46.78 & 0.5 & 1.6 & 0.5 & 11.5 & 12.0 & 21.5 & 14.0 & 24.5 & 45.5 & 1.5 & high \\
                 &      &      &       & 0.5 & 1.7 & 0.5 & 11.5 & 12.0 & 21.5 & 14.0 & 24.5 & 45.0 & 1.1 & low\\
  J132206-132354 & 4.71 & 9.48 & 46.18 & 0.5 & 1.4 & -0.3 & 12.0 & 12.4 & 20.1 & 15.3 & 24.5 & 46.5 & 1.9 & high \\
                 &      &      &       & 0.5 & 1.4 & -0.2 & 12.0 & 12.4 & 20.0 & 15.3 & 24.5 & 46.2 & 1.4 & low\\
  J132512+112329 & 4.42 & 9.60 & 46.89 & 0.5 & 1.5 & 0.65 & 11.0 & 12.7 & 20.6 & 15.0 & 25.0 & 46.3 & 0.1 & high \\
                 &      &      &       & 0.5 & 1.42 & 0.6 & 11.0 & 12.7 & 20.0 & 15.0 & 25.0 & 45.45 & 0.0 & low\\
  J134811+193520 & 4.4 & 9.18 & 46.48 & 0.5 & 1.6 & 0.5 & 11.5 & 12.0 & 22.0 & 14.0 & 24.5 & 46.2 & 1.8 & high \\
                 &      &      &       & 0.5 & 1.7 & 0.5 & 11.5 & 12.0 & 22.0 & 14.0 & 24.5 & 45.0 & 1.3 & low\\
  J141212+062408 & 4.47 & 8.78 & 46.78 & 0.5 & 1.75 & 0.5 & 11.0 & 12.7 & 20.2 & 15.0 & 25.0 & 45.85 & 0.8 & high \\
                 &      &      &       & 0.5 & 1.6 & 0.4 & 11.0 & 12.7 & 19.8 & 15.0 & 25.0 & 45.2 & 0.15 & low\\
  J142048+120545 & 4.02 & 9.18 & 46.48 & 0.5 & 1.7 & 0.5 & 11.5 & 12.0 & 21.5 & 14.0 & 24.5 & 46.0 & 1.5 & high \\
                 &      &      &       & 0.5 & 1.7 & 0.5 & 11.5 & 12.0 & 21.5 & 14.0 & 24.5 & 45.6 & 1.25 & low\\
  J143023+420450 & 4.72 & 9.18 & 46.78 & 0.5 & 1.5 & 0.4 & 11.5 & 11.8 & 21.0 & 14.0 & 24.5 & 47.0 & 1.8 & high \\
                 &      &      &       & 0.5 & 1.6 & 0.2 & 11.5 & 12.0 & 20.9 & 14.0 & 24.5 & 46.2 & 1.6 & low\\
  J145459+110927 & 4.93 & 9.18 & 46.11 & 0.5 & 1.5 & 0.4 & 11.5 & 11.8 & 21.5 & 14.0 & 24.5 & 46.0 & 1.8 & high \\
                 &      &      &       & 0.5 & 1.6 & 0.7 & 11.5 & 12.0 & 21.5 & 14.0 & 24.5 & 45.1 & 0.5 & low\\
  J151002+570243 & 4.31 & 9.36 & 46.40 & 0.5 & 1.5 & 0.25 & 11.5 & 11.8 & 21.2 & 14.0 & 24.5 & 46.8 & 1.8 & high \\
                 &      &      &       & 0.5 & 1.6 & 0.3 & 11.5 & 12.1 & 21.5 & 14.0 & 24.3 & 46.3 & 1.7 & low\\
  J153533+025419 & 4.39 & 9.0 & 46.48 & 0.5 & 1.7 & 0.22 & 11.5 & 11.8 & 21.0 & 16.0 & 27.5 & 46.2 & 1.9 & high \\
                 &      &      &       & 0.5 & 1.5 & 0.4 & 11.9 & 12.3 & 21.0 & 15.8 & 26.8 & 46.0 & 1.3 & low\\
  J162956+095959 & 5.0 & 8.95 & 46.30 & 0.5 & 1.45 & 0.22 & 12.0 & 12.0 & 21.0 & 16.0 & 27.5 & 46.4 & 0.7 & high \\
                 &      &      &       & 0.5 & 1.38 & 0.9 & 11.9 & 12.3 & 21.0 & 15.8 & 26.8 & 45.7 & -2.0 & low\\
  J164856+460341 & 5.36 & 9.0 & 46.65 & 0.5 & 1.45 & 0.22 & 12.0 & 12.0 & 21.0 & 16.0 & 27.5 & 46.7 & 0.7 & high \\
                 &      &      &       & 0.5 & 1.38 & 0.9 & 11.9 & 12.3 & 21.0 & 15.8 & 26.8 & 45.8 & -2.0 & low\\
  J165913+210115 & 4.784 & 9.40 & 46.49 & 0.5 & 1.45 & 0.5 & 12.0 & 12.0 & 21.0 & 16.0 & 27.5 & 46.5 & 0.75 & high \\
                 &      &      &       & 0.5 & 1.38 & 0.45 & 11.9 & 12.3 & 21.0 & 15.8 & 26.8 & 45.8 & 0.45 & low\\
  J170245+130104 & 5.466 & 9.0 & 46.30 & 0.5 & 1.55 & 0.4 & 12.0 & 12.0 & 20.5 & 16.0 & 27.5 & 46.6 & 1.2 & high \\
                 &      &      &       & 0.5 & 1.38 & 0.4 & 11.9 & 12.3 & 18.4 & 15.8 & 26.8 & 45.5 & 0.55 & low\\
  J171103+383016 & 4.0 & 8.81 & 46.26 & 0.5 & 1.6 & 0.22 & 12.0 & 12.0 & 21.0 & 16.0 & 27.5 & 46.5 & 0.7 & high \\
                 &      &      &       & 0.5 & 1.6 & 0.8 & 11.9 & 12.3 & 21.0 & 15.8 & 26.8 & 46.0 & -1.5 & low\\
  J171521+214547 & 4.01 & 8.48 & 45.90 & 0.5 & 1.4 & 0.22 & 11.8 & 11.8 & 21.0 & 15.0 & 27.5 & 46.8 & 1.0 & high \\
                 &      &      &       & 0.5 & 1.4 & 0.45 & 11.9 & 12.0 & 21.0 & 14.8 & 26.8 & 46.3 & 0.2 & low\\
  J172026+310431 & 4.67 & 9.88 & 46.60 & 0.5 & 1.45 & 0.22 & 12.0 & 12.0 & 21.0 & 16.0 & 27.5 & 46.4 & 1.0 & high \\
                 &      &      &       & 0.5 & 1.38 & 0.4 & 11.9 & 12.3 & 21.0 & 15.8 & 26.8 & 45.5 & 0.5 & low\\
  J195135+013442 & 4.11 & 9.18 & 46.30 & 0.5 & 1.7 & 0.22 & 11.7 & 11.7 & 21.0 & 16.0 & 27.5 & 46.9 & 1.3 & high \\
                 &      &      &       & 0.5 & 1.6 & 0.3 & 11.9 & 12.3 & 21.0 & 15.8 & 26.8 & 46.3 & 0.9 & low\\
  J213412-041909 & 4.334 & 9.26 & 46.74 & 0.5 & 1.7 & 0.4 & 12.0 & 12.0 & 21.0 & 16.0 & 27.5 & 46.9 & 0.6 & high \\
                 &      &      &       & 0.5 & 1.6 & 0.45 & 11.9 & 12.3 & 21.0 & 15.8 & 26.8 & 46.15 & 0.55 & low\\
  J222032+002537 & 4.2 & 9.30 & 46.40 & 0.5 & 1.7 & 0.35 & 12.0 & 12.0 & 21.0 & 16.0 & 27.5 & 46.7 & 0.7 & high \\
                 &      &      &       & 0.5 & 1.6 & 0.3 & 11.9 & 12.3 & 21.0 & 15.8 & 26.8 & 45.9 & 0.8 & low\\
  J231449+020146 & 4.11 & 9.00 & 46.54 & 0.5 & 1.6 & 0.35 & 12.0 & 12.0 & 21.0 & 16.0 & 27.5 & 46.7 & 0.7 & high \\
                 &      &      &       & 0.5 & 1.5 & 0.6 & 11.7 & 11.7 & 21.0 & 15.8 & 26.8 & 45.7 & 0.1 & low\\
  J235758+140205 & 4.35 & 9.00 & 46.48 & 0.5 & 1.6 & 0.22 & 12.0 & 12.0 & 21.0 & 16.0 & 27.5 & 46.7 & 0.7 & high \\
                 &      &      &       & 0.5 & 1.38 & 0.4 & 11.9 & 12.3 & 21.0 & 15.8 & 26.8 & 46.0 & 0.5 & low\\
\hline                                   
\end{tabular}
\end{sidewaystable*}
%

\section{VLA follow-up of the $z=5.793$ quasar J0502$-$34 in \cite{wolf24}}
\citet{wolf24} report the discovery of the $z=5.793$ quasar J050222$-$341158 (hereafter J0502$-$34).
This source was initially identified as a bright X-ray emitter in eRASS data processed with a preliminary version of the eROSITA reduction pipeline. 
In the most recent processing, however, the source is no longer detected according to the criterion defined in Section 2.2 of \citet{wolf24}.
This discrepancy is most likely due to random statistical fluctuations and the fact that J0502$-$34 lies close to the sensitivity limit of the eRASS survey.
The measured soft (0.2$-$2.3 keV observed frame) X-ray flux was 1.74$\times$10$^{-14}$ erg s$^{-1}$cm$^{-2}$.

Between the first and most recent pipeline releases—and prior to the publication of \citet{wolf24}—we conducted follow-up observations of J0502$-$34 with the Karl G. Jansky Very Large Array (VLA) to investigate its radio properties, such as radio spectral index and radio luminosity. 
No radio data for this source were available in the literature at that time.

The observations were carried out on 2023 March 26 in the L (1--2\,GHz) and S (2--4\,GHz) bands (program ID: VLA/23A-388, PI Belladitta) for a total of 1\,hr. 
The array was in the B-configuration, providing a maximum baseline of 11.1 km. In addition to the target source, the observing session included scans on the complex gain calibrator J0453$-$2808, and the flux density scale/bandpass calibrator 3C\,147.
The resulting on-target-source time was 10\,min per band. The observations utilized the 8-bit samplers of the VLA, and the WIDAR correlator was set up to deliver 16 sub-bands per receiver band. Each sub-band was 64\,MHz wide with 64 spectral channels at L-band, and 128\,MHz wide also with 64 spectral channels at S-band. The correlator integration time was 3\,s.

The data editing, calibration, and deconvolution were performed using the CASA software package \citep{mcmullin2007} version 6.4.1.12 and the CASA pipeline version 2022.2.0.64. Further manual editing was performed to flag frequencies heavily impacted by radio frequency interference. 

Self-calibration in phase was performed on the data of the target source using the continuum sources in the field to improve the fidelity of the images. At the conclusion of this process, and in order to minimize the impact of the strongest sources in the field, peeling of these sources was also performed.

The continuum images, one for each receiver band, were constructed using the CASA task tclean with the multi-term multi frequency synthesis deconvolution algorithm \citep{rau11} and setting nterms=2 to take into account variations of the spectral structure across the image. The Briggs weighting scheme with robust = 0.4 was chosen in tclean to balance angular resolution, rms noise, and sidelobe suppression.

\begin{figure*}
    \centering
    {\includegraphics[width=0.45\linewidth]{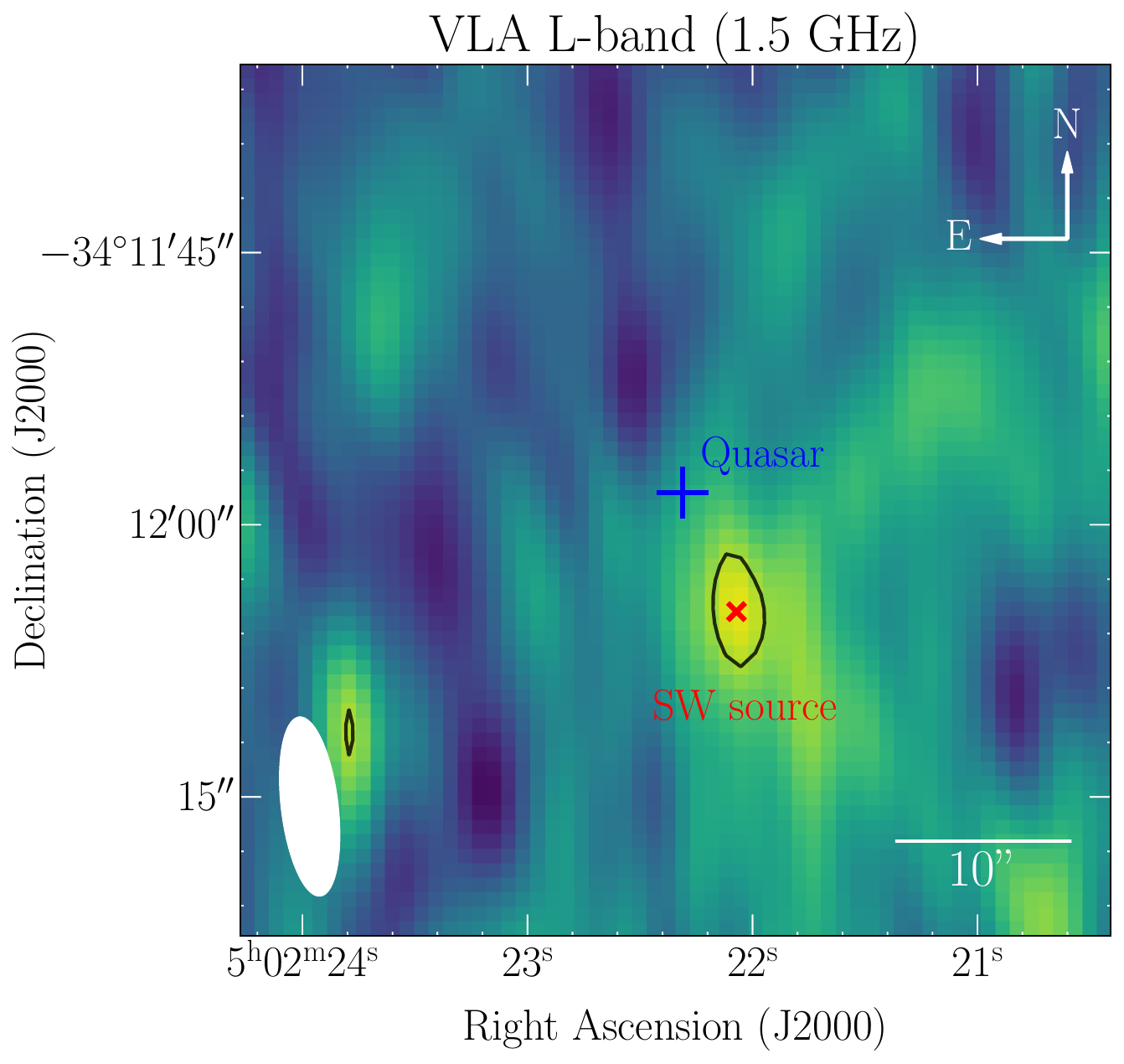}\hspace{0.1cm}
    \includegraphics[width=0.45\linewidth]{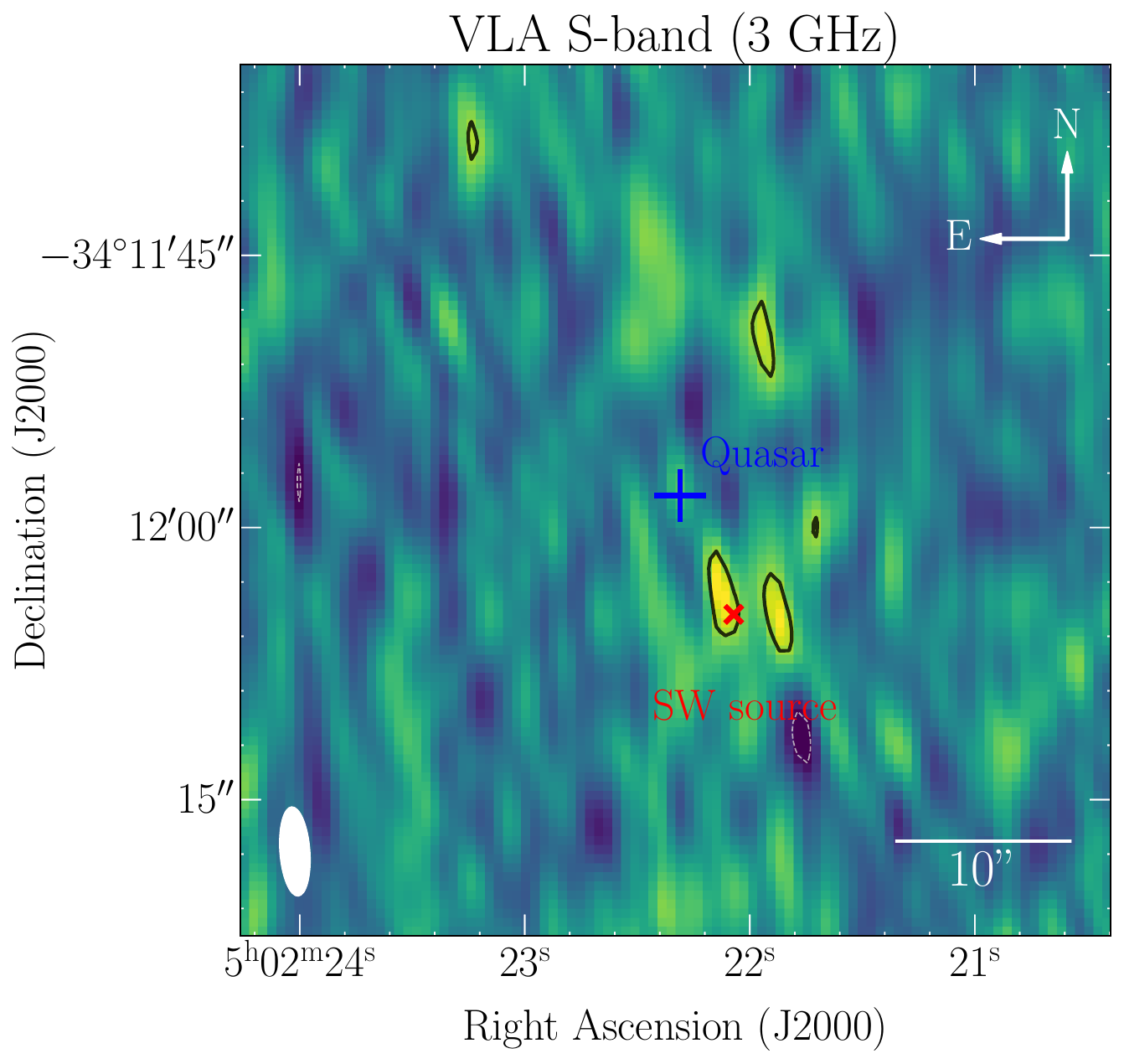}}
    \caption{VLA L and S-band radio maps of the field around the quasar J0502$-$34. Its optical position is shown as a blue cross, while the red X highlights the position of the optical SW source that coincides with a radio detection. Radio contours are plotted at 3$\sigma$ (RMS noise = 0.05 and 0.018 mJy\,beam$^{-1}$ for L and S-band, respectively). The synthesized beam of the radio observations are plotted on the bottom left corner: $9\farcs953 \times 3\farcs144$ PA=$6.781^\circ$ for L-band and $4\farcs949 \times 1\farcs671$ PA=$4.213^\circ$ for S-band.  North is up, East is left.}
    \label{fig:followup}
\end{figure*}

The resulting radio images (Fig.~\ref{fig:followup}) show that the detected radio emission is not spatially coincident with the quasar, but rather originates from a nearby source located toward the south-west.

The south-western radio counterpart has peak flux densities of 0.185$\pm$ 0.050\,mJy\,beam$^{-1}$ at L-band (3.7$\sigma$ detection) and 0.074 $\pm$ 0.018\, mJy\,beam$^{-1}$ at S-band (4.1$\sigma$ detection).
The derived spectral index between the two frequencies is $\alpha_{1.5}^{3.0} = 1.32\pm0.52$, indicative of a steep-spectrum radio object.
The position of the radio emission is consistent with an optical counterpart detected in all the five filters of Decals ($grizY$) suggesting a lower redshift nature of the source.

The absence of a radio detection at the position of J0502$-$34 supports the interpretation that it is not a jetted quasar, consistent with its non-detection in the eRASS X-ray data (see \citealt{wolf24} for details).

\end{document}